\pdfoutput=1
\documentclass[aps,pre, twocolumn, groupedaddress]{revtex4-1}
\usepackage{etoolbox} 
\usepackage{lipsum}
\usepackage{mathtools}
\usepackage{graphicx}
\usepackage{subcaption}
\usepackage{caption}
\usepackage{dcolumn}
\usepackage{amsmath}    
\usepackage{amssymb}
\usepackage{bm}
\usepackage{hyperref}
\usepackage{latexsym}
\usepackage{verbatim}
\usepackage[normalem]{ulem}
\usepackage{color}
\preprint{}
\begin{document}
\preprint{}
\title {Homing through Reinforcement Learning}

\author{Riya Singh$^{1}$}
\email{riyasingh.rs.phy24@itbhu.ac.in}
\author{Pratikshya Jena$^{1}$}
\email{pratikshyajena.rs.phy20@itbhu.ac.in }
\author{Anish Kumar$^{1}$}
\email{anishkumar.rs.phy22@itbhu.ac.in}
\author{Shradha Mishra$^{1}$}
\email{smishra.phy@itbhu.ac.in}
\affiliation{$^{1}$Department of Physics, Indian Institute of Technology (BHU), Varanasi, India 221005}
 
\date{\today}

\begin{abstract}
Homing and navigation are fundamental behaviors in biological systems that enable agents to reliably reach a target under uncertainty. We present a Reinforcement Learning (RL) framework to model adaptive homing in continuous two-dimensional domain. In this framework, the agent's state is given by its angular deviation from home, actions correspond to alignment or stochastic reorientation, and learning is driven by a radial-distance-based cost that penalizes motion away from the target, where the cost also acts as an effective signal guiding the agent towards the home. For a single self-propelled agent moving with constant speed, we find that the mean homing time 
$\langle T_{\mathrm{home}} \rangle$ exhibits a non-monotonic dependence on the rotational diffusion strength $D_r$, with an optimal noise level $D_r^\ast$, revealing a subtle interplay between exploration and goal-directed correction. Extending to two agents with soft repulsion, one agent consistently reaches home faster than the other, while in multi-agents system, repulsion ensures separation and the fastest agent becomes progressively faster as group size increases. Finally, we have compared the homing time obtained from the RL agent with that of an Active Brownian Particle (ABP) with resetting and a pure ABP (without resetting) under identical conditions. The RL-based agent consistently achieves shorter homing times with trajectories that are less noisy and more directed than both cases, while the pure ABP typically continues wandering near the target without reliable localization. Our results show that cost-driven learning, stochastic reorientation, and inter-agent interactions enable efficient adaptive navigation, linking individual and collective homing. This reinforcement learning framework captures key biological features such as feedback-based route learning, randomness to escape unfavorable orientations, and mutual 
coordination.
\end{abstract}
\maketitle

\section{Introduction}
 Homing - the ability of organisms to reliably return to specific locations such as nests, burrows, or dens, is a canonical example of goal-directed spatial navigation in complex and uncertain environments. Across species, this behavior is realized through diverse mechanisms, including path integration in desert ants \cite{wehner2003desert}, the combined use of geomagnetic cues and visual landmarks in pigeons \cite{biro2007pigeons}, and echolocation-based navigation in bats \cite{diebold2020adaptive}, underscoring its evolutionary importance for efficient foraging, reproduction, and survival. In recent years, the study of homing has extended beyond biology, attracting growing interest from robotics, physics, and control theory. Motivated by the work of Paramanick \textit{et al.}~\cite{paramanick2024uncovering}, which demonstrated that foraging robots exhibit universal statistical features in their homing trajectories largely independent of specific design details, we explore the fundamental principles governing adaptive navigation. Additionally, homing has been studied using computational and theoretical models \cite{benichou2005optimal,bartumeus2005animal,painter2015navigating}, as well as stochastic resetting strategies ~\cite{evans2011diffusion,evans2020stochastic,evans2013optimal,christou2015diffusion,pal2019first,durang2019first} that enhance the search efficiency. 
 
 Despite these advances, existing approaches are often limited in their ability to capture adaptive decision-making in uncertain environments: theoretical models rely on predefined navigation rules \cite{shen2024comparative,codling2008random} , experiments are limited by biological variability \cite{sumpter2010collective} , and simulations typically assume fixed stochastic dynamics \cite{bartumeus2008fractal}, lacking the flexibility to include learning-driven adaptation. 
 
 Reinforcement Learning (RL) provides a natural framework to overcome these limitations by casting homing as a sequential decision-making problem
\cite{botvinick2019reinforcement,mnih2015human}. 
Unlike supervised learning, which relies on labeled data, or unsupervised learning, which identifies hidden patterns within datasets, RL involves an agent that learns by performing actions, observing their outcomes, and receiving feedback from the environment in the form of cost/reward 
 \cite{sutton1998reinforcement,ghasemi2024introduction}.  And this form works for  single as well as  multi-agent systems \cite{leibo2017multi,yang2020overview,wang2022model,yang2025game,zhang2021multi,nowe2012game,tuyls2005evolutionary} where cooperation and competition emerge, making it a powerful framework to study collective homing.

In the present work, we utilized  the RL method for  active or self-propelling agents/particles.  These self-driven particles  consume energy to produce directed motion and inherently they are  far from equilibrium \cite{marchetti2013hydrodynamics,semwal2024dynamics, jena2024polarised,pincce2016disorder, wolf2011janus, buttinoni2013dynamical,brambilla2013swarm}. 
Within this context, we develop a Q-learning–based RL model that enables a self-propelled agent moving with constant speed to efficiently reach its home location. Among other RL methods—such as SARSA (State-Action-Reward-State-Action) \cite{humayoo2024segmenting,mohaniterative}, Proximal Policy Optimization \cite{schulman2017proximal,wu2021coordinated}, and actor–critic algorithms \cite{konda1999actor,oren2024value} — Q-learning \cite{jia2023q,watkins1992q} provides a natural and convenient framework for the homing task, as alternative methods often involve on-policy updates or optimized parameterized policies that add unnecessary complexity, whereas Q-learning is simple, fully model-free, and directly updates action values. 

Our primary objective in this study is to investigate a minimal stochastic optimization framework for homing. We therefore begin by first analyzing the navigation behavior of a single active agent. A central quantity in our analysis is the homing time, $T_{\rm home}$, which reflects how efficiently an agent returns to its home. In real organisms, $T_{\rm home}$ varies with age, climatic conditions, and physiological state which regulate the stochasticity of motion. In our model, this variability is captured by the rotational diffusion strength $D_r$, which controls angular noise and resetting frequency, allowing us to mimic a broad spectrum of biological and physical influences on homing efficiency. \\
For comparison, we also analyze the corresponding dynamics of an Active Brownian Particle (ABP) subjected to the same angular noise. Across a wide range of \(D_r\), the RL agent consistently achieves shorter homing times and less noisy trajectories than the ABP. For completeness, we also compare with a pure ABP (without resetting), where the particle undergoes persistent motion with stochastic reorientation due to rotational diffusion, which allows space exploration but
does not ensure convergence to the target.
Accompanied by distinct dynamical signatures, the resetting statistics show a monotonic increase in both the mean number of resetting events and the effective resetting rate with respect to $D_r$.

\begin{figure*}
    \centering
    \includegraphics[width=1.0\linewidth]{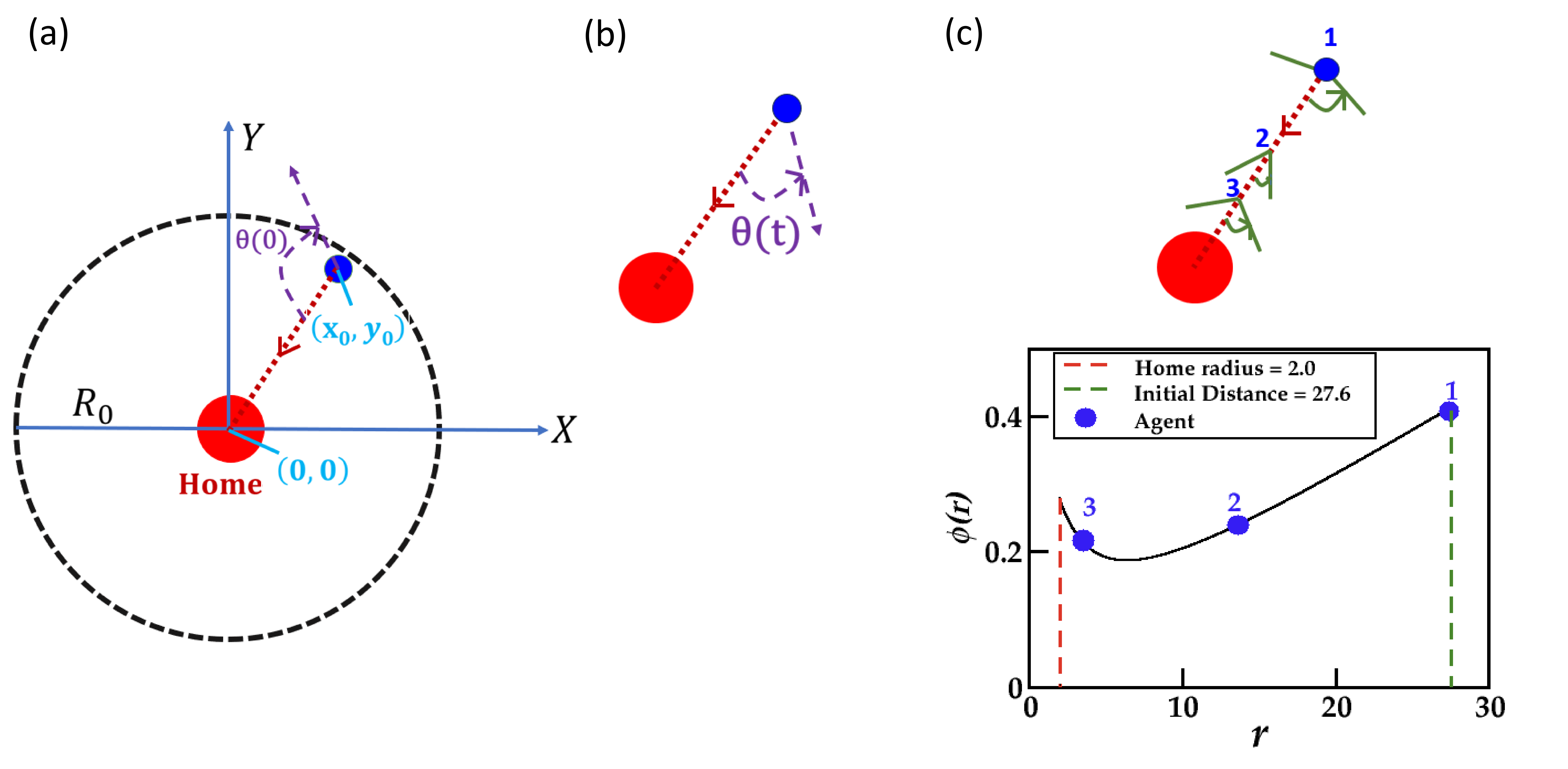}
    \caption{(a) (Color online) Schematic representation of the homing model showing the circular domain of radius $R_0$. The agent (indicated by filled blue circle) starts from an initial position at $(x_0, y_0)$ in the $xy$-plane, with its initial orientation $\theta(0)$ (shown by purple arrowed dashed line) measured from homing direction taken as reference axis (shown by red arrowed dotted line) from agent towards home (marked by red filled circle at center of domain).
    (b) Shows the agent’s instantaneous orientation as measured from home direction as denoted by $\theta(t)$. (c) The angular threshold $\phi$ is shown at three representative positions — (1) the initial point, (2) midway along the trajectory, and (3) near the home region corresponding to same three positions marked on plot. The solid green line represents the evolution of the threshold angle $\phi$ about the home direction, which starts at a large value, decreases to a minimum as the agent aligns towards home, and slightly increases again upon approaching the home location as can be seen in plot also which shows the variation of $\phi(r)$ with distance from the home location. The function is evaluated starting from the initial position $(19, 20)$, corresponding to an initial distance, $r_0 = 27.6$ from home, and continues up to the home region at $r = 2.0$. Thus, the angular threshold gradually decreases as the agent approaches the home position, enforcing greater directional precision in the vicinity of the home while maintaining sufficient freedom for exploration near home.}
   \label{fig:model}
\end{figure*}

Building on the single-agent dynamics, we extend the model to two-agent and multi-agent systems, where trajectories are coupled through short-range repulsive interactions that prevent overlap and mimic avoidance behavior. We identify three distinct regimes: in the single-agent case, the mean homing time $\langle T_{\mathrm{home}} \rangle$ exhibits a non-monotonic dependence on the rotational diffusion strength $D_r$, indicating the presence of an optimal noise level $D_r^\ast$; for two interacting agents, asymmetry emerges, with one agent consistently reaching home faster; and in multi-agent populations, repulsive interactions maintain separation while the fastest agent increasingly outperforms the others as the group size grows.

Together, these regimes reveal general principles of learned navigation that extend from a single agent to interacting groups, offering guidance for both biological studies and the design of engineered active systems. The emergence of faster homing agents in larger groups suggests strategies to improve transport efficiency \cite{lai2024reinforcement}, reduce search times \cite{munoz2025learning}, and improve coordinated delivery in both robotic \cite{kormushev2013reinforcement} and biomedical applications \cite{niazmand2024deep}.

The remainder of this paper is organized as follows. In section ~\ref{2}, we describe the details of the model, the Q-learning framework, and the state–action definitions. Section ~\ref{3} presents the results: single-agent dynamics is analyzed in section ~\ref{3.1}. The evolution of the Q-matrix elements and the associated action-selection dynamics are analyzed in section ~\ref{3.2} followed by two-agents and multi agents extensions in section ~\ref{3.3} and section ~\ref{3.4}, respectively. A direct comparison between active Brownian particle (ABP) with resetting and RL agent is presented in section ~\ref{3.5}, while the behavior of a pure ABP (without resetting) is discussed in section~ \ref{3.6}. Finally, the conclusions are summarized in section ~\ref{4}.\\

\section{Model}{\label{2}}

In our model, homing occurs in a continuous two-dimensional circular domain, reflecting the inherently continuous nature of animal and robotic navigation and avoiding lattice artifacts \cite{dhar2022robust,dhar2025adaptive}.
The spatial extent of the arena is defined as shown in Fig. \ref{fig:model}(a) by an effective circular domain of radius $R_{0}$ (indicated by the black dashed circle) centered at the origin $(0,0)$ with reflecting boundary conditions which prevents unbounded escape of the particle and provides a minimal representation of navigation in finite domain, such that agent remains within the domain and continues to explore the environment until it reaches the target
\cite{chepizhko2021revisiting}. The agent (shown by a filled blue circle) starts at an initial position $(x_0, y_0)$ inside the domain, away from both the boundary and the home. The home (marked by a filled red circle) is defined as a circular target region of radius $r_{\mathrm{home}} = 2$ centered at the origin. 

Consistent with this setup, at each simulation step \( t \to t + \Delta t \), the agent computes its instantaneous radial distance from the home location as:
\begin{equation}
r(t) = \sqrt{x^2(t) + y^2(t)}.
\label{eq:radial_distance}
\end{equation}

where, ${\it {\bf r}(t)} = \big(x(t),\, y(t)\big)$ denotes the position of agent at time $t$. This radial distance determines the proximity of the agent to the home and serves as the principal spatial variable governing both the cost evaluation and the angular threshold as discussed below.

The agent’s orientation is characterized by an angular deviation \(\theta(t)\), defined as the minimal signed angle between the agent’s heading direction (shown by the purple arrowed dashed line) and the instantaneous direction toward the home (indicated by red arrowed dotted line), which is taken as the reference axis as illustrated in Fig. \ref{fig:model}(b). By construction, $\theta(t) = 0$ corresponds to perfect alignment toward the home, while positive (negative) values indicate clockwise (counterclockwise) deviations. The initial orientation of agent $\theta(0) \in [-\pi, \pi)$ is chosen randomly, reflecting the absence of prior directional information. To maintain consistency in orientation dynamics, this deviation is confined to the full interval \([-\pi, \pi]\) rather than \([-\pi/2, \pi/2]\). This modeling choice allows the agent to execute both minor steering corrections and large-angle 
reorientation, including near-reversals, which are essential in biological and robotic navigation when an agent loses track of its direction or encounters environmental perturbations 
\cite{couzin2005effective,hein2016natural}.

To regulate orientation precision, we introduce an angular threshold, \(\phi(r(t))\) (shown by green solid lines about the homing direction) as illustrated in Fig. \ref{fig:model}(c), which specifies the maximum allowable deviation in the agent’s turning angle, \(\theta(t)\). All angles are rescaled by $\pi$.
 At each time step, this threshold constrains the allowed reorientation according to:

\begin{equation}
\phi(r(t)) = \frac{\pi}{2}\left( \frac{r(t)}{R_0} \right) + \tan^{-1}\!\left( \frac{r_{home}}{r(t)} \right),
\label{eq:threshold}
\end{equation}

where \(R_0\) is the radius of the circular domain and \(r_{home}\) represents the radius of home region. 
In natural navigation processes such as chemotaxis and homing,
organisms typically exhibit stronger reorientation and a greater directional response near the target
to ensure accurate localization \cite{tu2008modeling,celani2010bacterial}. Motivated by this behavior, the two terms in  $\phi(r(t))$ serve distinct and complementary roles. The first term introduces a radial dependence that decreases linearly with distance as the agent moves closer to home, thereby decreasing the directional constraint at larger distance from home and promoting exploratory motion when the particle is far from the target, allowing it to effectively search and move towards the home from distant regions. However, a purely linear form leads to a continuously decreasing (monotonic) angular threshold as the particle approaches the target. In this situation, the condition $|\theta(t)| \le \phi(r(t))$ as shown below in Eq. \ref{eq: state_description}, becomes increasingly difficult to satisfy at small distances. As a result, the particle tends to remain in the misaligned state and continues to move around the target region without successfully settling into it.
To overcome this issue, we include the second term $\tan^{-1}(r_{home}/r(t))$, which  ensures that $\phi(r(t))$ does not keep decreasing near the target and instead maintains sufficient angular tolerance for reorientation near the home  as shown in plot of Fig. \ref{fig:model}(c). As a consequence, the chosen form of $\phi(r(t))$ develops a minimum at an intermediate distance, which plays an important functional role in the dynamics. The presence of the minimum ensures a controlled spatial balance between exploration at intermediate distances and reliable alignment near the target, a feature that cannot be achieved with a purely monotonic function. Consequently, the particle converges reliably towards the home, thereby avoiding instabilities in its vicinity.

\begin{figure*}
    \centering
    \includegraphics[width=1.0\linewidth]{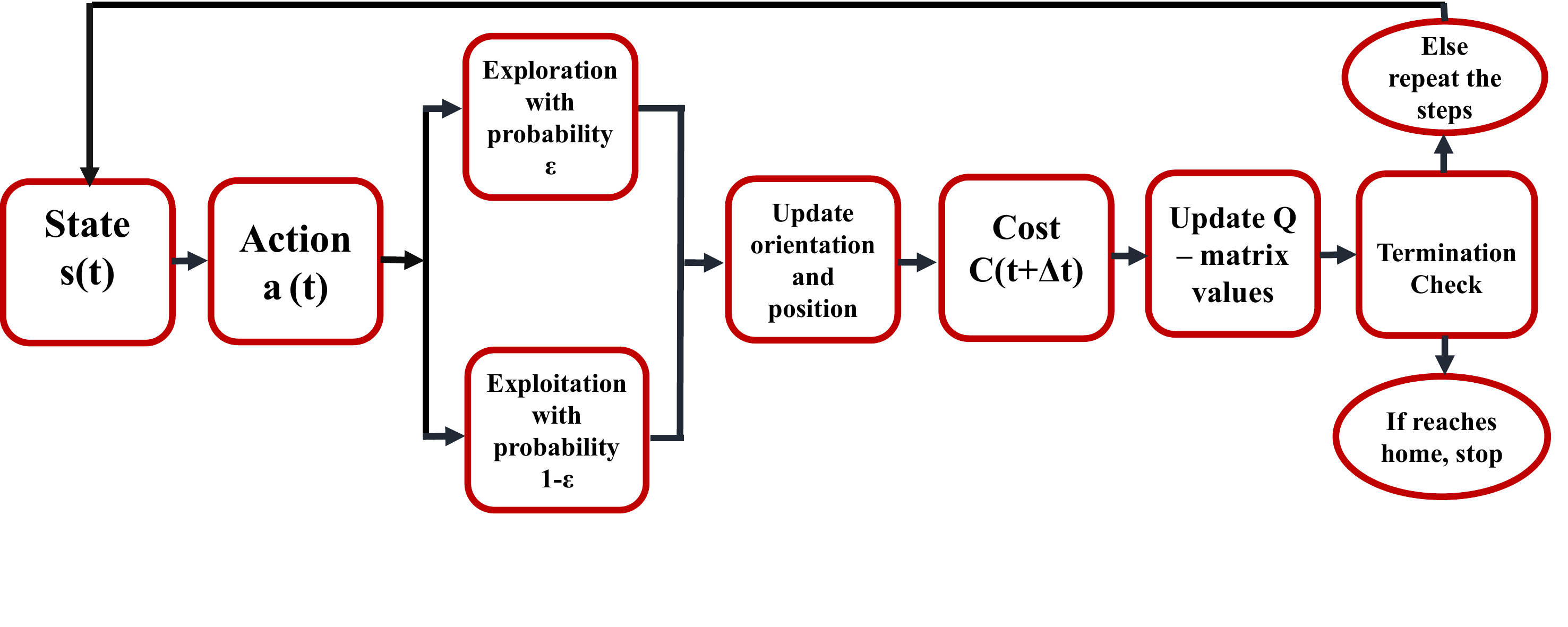}
    \caption{(Color online) Flowchart illustrating the reinforcement learning (RL) framework used for homing. The algorithm starts from the current state and at each iteration, an action is selected using an $\varepsilon$-greedy exploration--exploitation policy, followed by updates of the agent’s orientation and position. The cost is then evaluated based on the distance from the home location, and the Q-values are updated accordingly. A check is performed to determine whether the agent has reached the home; if not, the procedure is repeated from the updated state, otherwise the algorithm terminates.}

    \label{fig:RL_flowchart}
\end{figure*}

Based on the computed angular deviation, \(\theta(t)\) and angular threshold, \(\phi(r(t))\) as in Eq. \ref{eq:threshold} , the agent`s state is discretized into two classes, thereby minimizing the complexity of the state space while preserving essential decision information. 
The state variable \(s(t)\) is defined as:
\begin{equation}
s(t) =
\begin{cases}
1, & \text{if } |\theta(t)| > \phi(r(t)), \\[4pt]
2, & \text{if } |\theta(t)| \le \phi(r(t)).
\end{cases}
\label{eq: state_description}
\end{equation}
The use of the absolute deviation ensures that only the magnitude of the misalignment determines the state, since clockwise and counterclockwise deviations are equally detrimental from a homing perspective. Physically, these two states represent distinct behavioral modes:
state $s(t) = 1$ represents a misaligned configuration in which the agent’s heading exceeds the permitted angular threshold and thus requires corrective 
reorientation to avoid inefficient wandering, whereas state $s(t) = 2$ denotes an aligned configuration in which the agent remains within the angular threshold and needs only minor adjustments to continue progressing towards home.

Such binary state formulations are commonly employed in active search and foraging models, where orientational resets are observed to accelerate convergence and minimize the mean first-passage time 
\cite{evans2011diffusion,evans2020stochastic,pal2019first}.

Given its current state, the agent selects an action \(a(t) \in \{1, 2\}\) using an 
\(\varepsilon\)-greedy policy:
\begin{equation}
a(t) =
\begin{cases}
\text{random action}, & \text{with probability } \varepsilon,\\[6pt]
\displaystyle \mathit{arg\,min}\; Q[s(t),a(t)], & \text{with probability } 1-\varepsilon.
\end{cases}
\label{eq: action_RL}
\end{equation}
here, the \textit{arg\,min} operator returns the action \(a(t)\) that yields the minimum Q-value among all available actions in state \(s(t)\). The \(\varepsilon\)-greedy strategy promotes a balance between exploration and exploitation \cite{reddy2016infomax}: with 
probability \(\varepsilon\), the agent explores by selecting a random action, while with probability 
\(1 - \varepsilon\), it exploits its learned knowledge by choosing the action that minimizes the expected cost. Once an action is chosen, the agent updates its orientation accordingly:
\begin{equation}
\theta{(t+\Delta t)} =
\begin{cases}
0, & a(t) = 1, \\[6pt]
\theta(t) + \sqrt{2 D_r \Delta t}\,\zeta, & a(t) = 2 .
\end{cases}
\label{eq: action_update}
\end{equation}
\noindent
where \(D_r\) is the rotational diffusion strength and \(\zeta\) is a uniformly distributed random variable in the interval \([-\pi/2, \pi/2]\). The bounded range ensures that exploration remains symmetric and moderate, avoiding unrealistic large angle fluctuations within a single update. Both $\theta(t)$ and $\theta{(t+\Delta t)}$ are defined with respect to the instantaneous home direction, and $\theta{(t+\Delta t)} = 0$ denotes perfect alignment towards home. The action $a(t)=1$ represents a situation in which the particle actively corrects its orientation and aligns itself with the target after a phase of exploratory motion. In this sense, the reset of $\theta(t+\Delta t)$ to zero should be viewed as a minimal and effective description of such reorientation processes, where instead of modeling a gradual turning, the system instantaneously aligns with the target direction. In contrast, the exploratory action ($a(t)=2$) represents stochastic angular fluctuations corresponding to situations where the signal is weak relative to noise and the agent explores different directions through random reorientation. Such behavior is commonly observed in run and tumble dynamics of bacteria, where motion consists of alternating phases of exploration (runs) and sudden reorientation events (tumbles), during which the direction of motion is rapidly changed \cite{semwal2024dynamics,zamora2024exploring}. Similarly, in animal navigation and chemotaxis, organisms often perform intermittent corrections to align with a target or gradient.

After updating orientation, the agent’s position is advanced using a kinematic rule:

\begin{equation}
\mathbf{r}{(t+\Delta t)} = \mathbf{r}{(t)} + v_{0}\Delta t\, \hat{\mathbf{n}}{(t+\Delta t)},
\label{eq: position_update}
\end{equation}

where $\mathbf{r}{(t)}=(x(t),\, y(t))$ and 
$\hat{\mathbf{n}}{(t+\Delta t)}=(\cos\theta{(t+\Delta t)},\, \sin\theta{(t+\Delta t)})$
is the unit vector in the direction of motion, \(v_0 \) is the agent’s constant speed and \(\Delta t\) is the discrete time step. 

To evaluate the effectiveness of each action within this evolution, we define the cost function, \( C{(t+\Delta t)} \) guiding the learning process based on the change in radial distance resulting from the chosen action.
The instantaneous cost at time $t+\Delta t$ is defined as:
\begin{equation}
C(t+\Delta t) = |\mathbf{r}(t+\Delta t)| - |\mathbf{r}(t)|,
\label{eq: cost_function}
\end{equation}
where the cost $C(t+\Delta t)$ is the radial displacement of the agent.
This definition ensures that any movement towards the home lowers the cost, thereby providing a simple and continuous measure of progress towards the target. At the same time, the cost acts as an effective
signal field that guides the agent through gradient-like information. We choose a smooth long-ranged form to avoid introducing additional parameters such as sensing cutoffs and to isolate the role of stochasticity and learning in navigation. Because the cost depends only on the change in radial distance, and not on the specific state or action taken, minimizing the cost through Q-value updates thus naturally promotes convergence toward the home location. Since the model consists of two discrete states and two possible actions. Accordingly, the agent  is associated with a 2×2 Q-matrix, whose elements—referred to as Q-values—quantify the expected cost of taking a given action in a given state. 

The Q-matrix is updated at every time step. Initially, all Q-values are set to zero, and the Q-value associated with the current state–action pair \([s(t), a(t)]\) is updated after each step according to the following equation:
\begin{equation}
Q[s(t), a(t)] \leftarrow (1 - \alpha)\, Q[s(t), a(t)] + \alpha\,[C{(t+\Delta t)}],
\label{eq: Q_update}
\end{equation}
where \(\alpha\) is the learning rate that controls the rate of adaptation. The symbol '$\leftarrow$' denotes an update operation, indicating that the
quantity on the left-hand side is reassigned the new value computed on the right-hand side. 
This choice corresponds to a simplified version of the Q-learning framework in which the update is governed primarily by the immediate outcome of the action.
Such a formulation is appropriate here because reducing the instantaneous distance lowers the cost and is directly aligned with minimizing the total time required to reach the target. Similar immediate-cost-based reinforcement
learning frameworks have also been widely used in earlier studies of stochastic optimization and adaptive dynamical processes \cite{kumar2025adaptive, pramanik2025run, sampat2022ordering, durve2020learning}. For ease of comprehension and quick visualization of the methodology, the overall process is summarized in the flowchart as shown in Fig. \ref{fig:RL_flowchart}.
Over successive iterations, RL mechanism enables the agent to discover an optimal sequence of reorientations that minimize cumulative cost, thereby facilitating reliable homing even in the presence of environmental noise.\\
In the simulation, we have used agent's constant speed \(v_0 = 0.1\), time step \(\Delta t = 0.05\), learning rate \(\alpha = 0.001\), and exploration probability \(\varepsilon = 0.3\). This choice of \(\alpha\) and \(\varepsilon\) achieves a suitable balance between exploration and convergence stability, consistent with previous RL studies \cite{kumar2025adaptive,pramanik2025run}. Extremely large \(\alpha\) can cause unstable or divergent Q-value updates, while excessively small values slow convergence and reduce adaptability. Similarly, a moderate \(\varepsilon\) ensures sufficient stochastic exploration without overwhelming the learned policy. The rotational diffusion strength \(D_r\) is varied from \(0.5\) to \(20\). The arena has radius \(R_0 = 35\), and the agent is initially positioned at distance  $r_0 \sim 27$ from the center of the home. This choice avoids trivial cases such as immediate homing from the center or boundary-driven motion due to reflections and provides a meaningful initial condition. Each agent is assigned a finite size with radius $r_p = 0.35$. As an illustrative mapping to a biological system, one may interpret the agent as a bacterium of characteristic size $r_p \sim 1\,\mu{\rm m}$ swimming at a speed $v_0 \sim 20$--$30\,\mu{\rm m/s}$, values typical of \textit{E. coli} reported in experiments \cite{bhattacharjee2019bacterial,mathijssen2019oscillatory}. \\
\begin{figure*}[htbp]
    \centering
\includegraphics[width=0.8\linewidth]{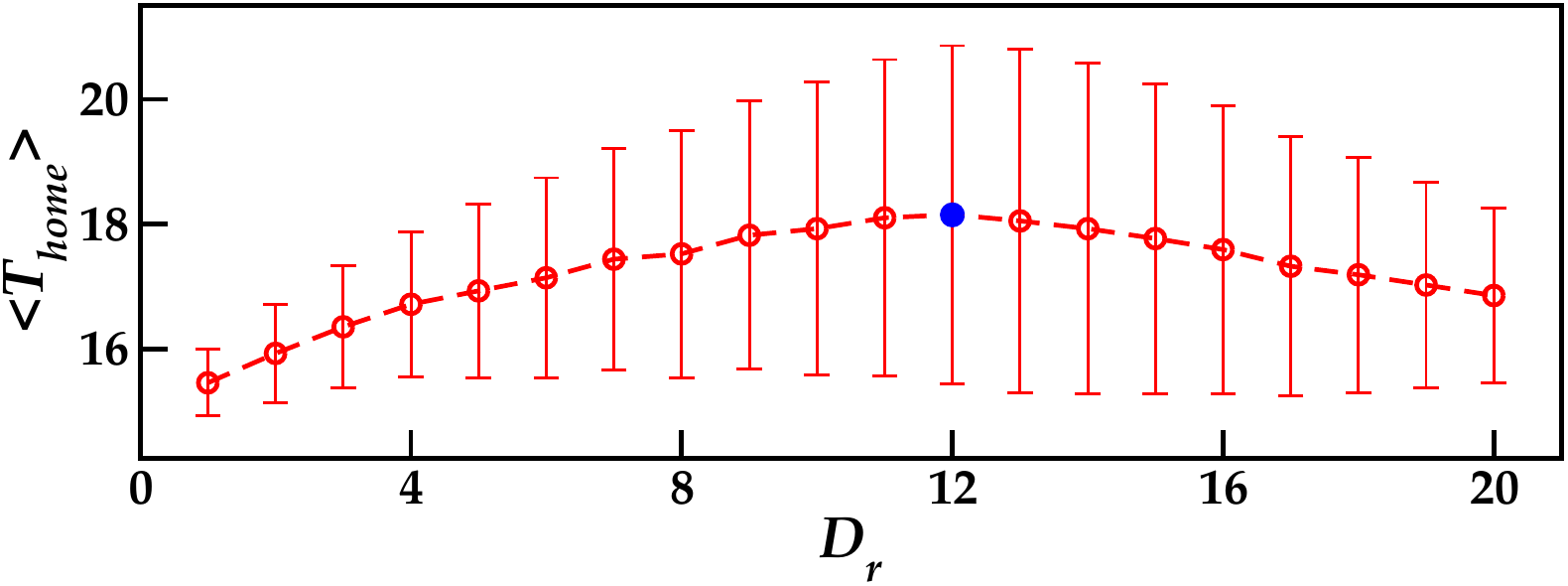}
    \caption{ (Color online) Mean homing time, $\langle T_{\text{home}} \rangle$ $vs.$ rotational diffusion strength, $D_r$ 
    for $\varepsilon = 0.3$ and $\alpha = 0.001$, averaged over 1500 independent realizations. The mean homing time, $\langle T_{\text{home}} \rangle$ initially increases with $D_{r}$, then becomes nearly constant over an intermediate range, and finally beyond an optimal value $D_r^\ast \sim$ 12 (blue), the homing time begins to decrease as $D_r$ is further increased. 
    The symbols show the data points and  the dashed line is a guide to the eye while error bars indicate the corresponding standard deviations of the homing-time distribution obtained from independent trajectories.}
    \label{fig:mean_thome_vs_Dr}
\end{figure*}
The confinement radius $R_{0}$ serves as the characteristic length scale of the system  and represents the spatial extent over which the agent explores before reaching the home region.
A characteristic time scale is defined as the time required for the agent to traverse the domain radius \(R_0\) with its  speed \(v_0\), given by
$T_0 = \frac{R_0}{v_0} \Delta t$.
\label{eq: Intrinsic_time}
All temporal quantities in the analysis are rescaled by this characteristic time scale. Thus, the evolution time characterizes the duration required for the agent to learn and execute a successful homing strategy relative to the time needed to cross the system ballistically along a straight trajectory without reorientation. For better statistics, unless stated otherwise, all the results are averaged over 1500 independent realizations. Building upon the single-agent Q-learning homing model, we extend the framework to two-agents and multi-agents system by introducing interactions between agents. A full description of the form of interaction potential, initial conditions and size of agents, parameter choices, and implementation details is provided in the Appendix \ref{Dynamics:two_multi}.

\begin{figure*}[htbp]
    \centering
    \includegraphics[width=1.0\textwidth]{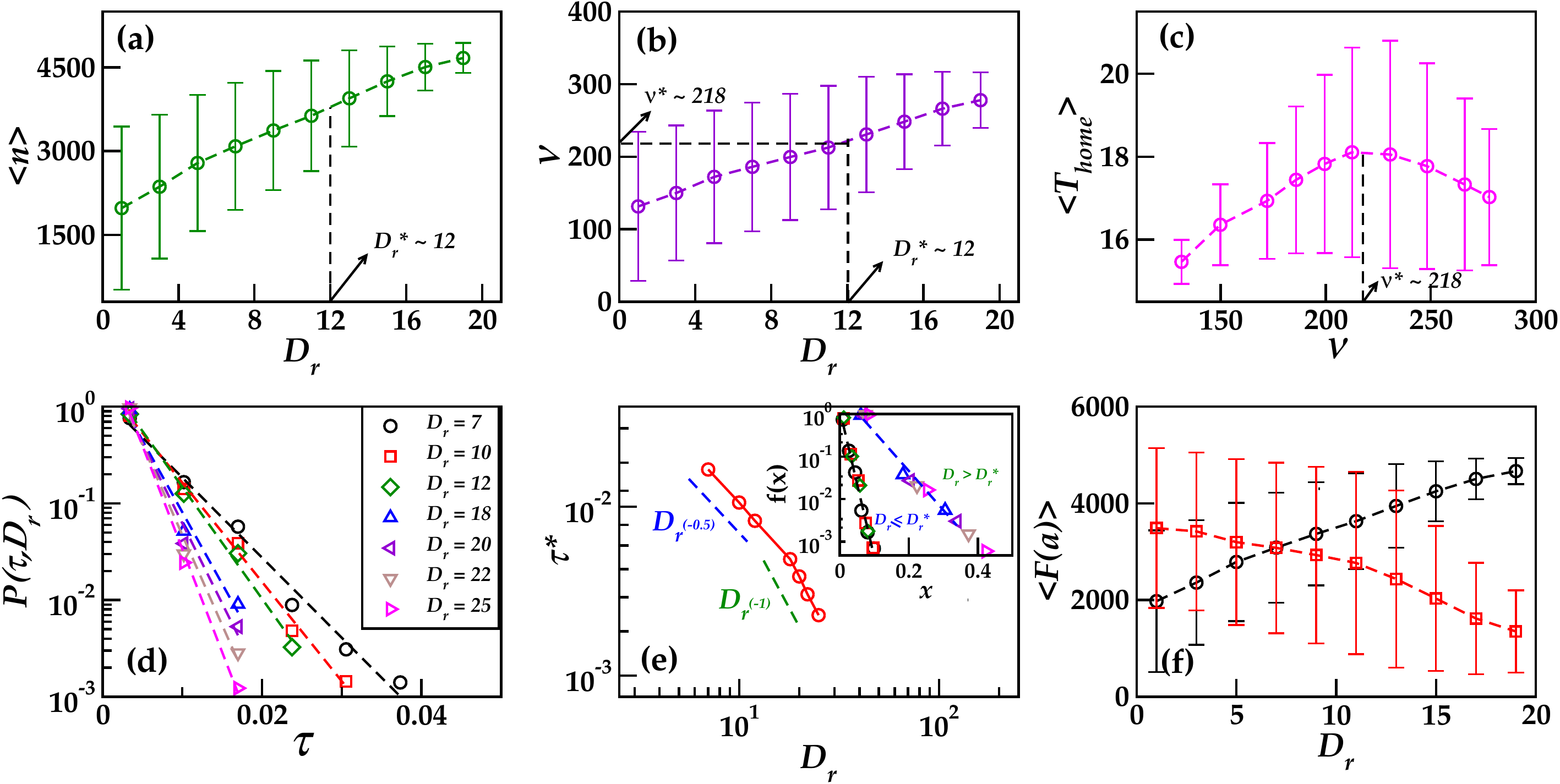}\hfill
   
    \caption{
    (a) (Color online) Mean number of resettings $\langle n\rangle$ $vs.$ $D_r$, averaged over 1500 independent realizations. 
    (b) Frequency of resetting $\nu = \left\langle \frac{n}{T_{\rm home}} \right\rangle$ $vs.$ $D_r$
    (c) Mean homing time $\langle T_{\mathrm{home}} \rangle$ vs reset frequency $\nu$. 
    (d) Probability distribution $P(\tau,D_r)$ of time intervals between consecutive resets. 
    (e) The characteristic time $\tau^{*}$ as a function of $D_r$ shows a power-law dependence with two distinct scaling regimes. 
   For $D_r \leq D_r^{*}$, $\tau^{*} \sim D_r^{-\alpha}$ with $\alpha = 0.5$, while for $D_r > D_r^{*}$, $\tau^{*} \sim D_r^{-\alpha}$ with $\alpha = 1$. The inset shows the collapsed plot of panel (d) using the scaling function $f(x)$ vs. $x$ for   $D_r \leq D_r^{*}$ and  $D_r > D_r^{*}$.
 (f) Shows the effect of rotational noise $D_r$ on frequency of action-selection statistics where black circles and red squares correspond to number of times action 1 and action 2 are taken respectively, averaged over time and realizations. Increasing noise drives a crossover in the learned policy, characterized by an enhanced preference for action~1 and a systematic reduction in the selection of action~2. Error bars show the standard deviation in respective quantities rather than the statistical uncertainty of the ensemble-averaged quantities in each plot.}
    
    \label{fig:resetting_analysis}
\end{figure*}
\section{Results}{\label{3}}
\subsection{Single-agent Regime}{\label{3.1}}

We first focus on the dynamics of a single agent navigating towards home. In our simulation, we explore the effects of the exploration and learning parameter on the agent's homing performance. Initially, the exploration probability is fixed at $\varepsilon = 0.3$ and the learning rate at $\alpha = 0.001$ to evaluate the agent’s homing performance under varying values of $D_r$. For each $D_r$ value, the mean homing time $\langle T_{\mathrm{home}} \rangle$ is computed.

$T_{\mathrm{home}}$ is the time required for the agent to reach a target region of radius $r(t) \le 2.0$ around the origin, measured in simulation steps and converted to actual time using $\Delta t$. The mean $\langle T_{\mathrm{home}} \rangle$ is obtained by averaging over 1500 independent realizations starting from the different  initial position and orientation. The initial distance of the agent from the home is chosen in the range $r_0 \sim 27$.

$\langle T_{\text{home}} \rangle$ exhibits a non-monotonic dependence on $D_{r}$, it first increases with $D_{r}$, then becomes nearly constant over an intermediate range, and finally beyond an optimal value $D_r^\ast \sim 12$, the homing time begins to decrease as $D_r$ is further increased as shown in Fig. \ref{fig:mean_thome_vs_Dr}. After establishing this baseline, we further investigated the variation of  $\langle T_{\mathrm{home}} \rangle$ against $D_r$ for different ($\varepsilon$,$\alpha$) combinations (see Fig. \ref{fig:Thome_vs_Dr_epsilon_alpha} in Appendix \ref{Parameters} for details).

The bars associated with $\langle T_{\mathrm{home}} \rangle$ represent the standard deviation of the
homing-time distribution obtained from independent trajectories and are intended to quantify the spread of homing times rather than the statistical uncertainty of the ensemble-averaged homing time. Similar to the mean homing time, the standard deviation also exhibits a non-monotonic dependence on $D_r$. For small $D_{r}$, the standard deviation is small and increases with $D_{r}$, reaching a maximum within an optimal range of $D_{r}$, and then decreases again at higher noise levels. At low noise, trajectories are nearly deterministic, yielding similar homing times and small variability. At intermediate $D_r$, noise competes with learning, producing a wide spread of paths and maximal fluctuations, 
while at high $D_r$ frequent reorientation leads to statistically similar, diffusive trajectories and reduced variability. To verify that the observed standard deviation reflect intrinsic properties of the system and are not reducible by averaging over many realizations, we compute the homing time over different independent realizations. (see Fig. \ref{fig:Thome_errorbars_ensembles} in Appendix \ref{Error_bar} for details). The standard deviations remain essentially unchanged, confirming that they are an intrinsic property of the system itself rather than a statistical artifact. Furthermore, the mean homing times obtained using larger numbers of realizations remain within the $95\%$ confidence interval of the $N_{\rm ens}=1500$ dataset, demonstrating that $1500$ realizations are sufficient to obtain statistically converged estimates, as discussed in detail in Appendix \ref{Error_bar}.

We now examine the effect of random resetting on the homing time.
In previous studies it is observed that the random resetting can substantially optimize search processes by making the mean first-passage time (MFPT) finite and often minimizing it at an optimal reset rate, thereby improving the efficiency of locating targets compared to search without resetting \cite{bressloff2020modeling}.
To quantify how random resetting affects the homing process, we measure the mean number of resetting events, $\langle n \rangle$ that occur during homing for different values of $D_r$. In the simulation, a resetting event corresponds to the agent’s complete reorientation towards home direction, implemented when $a(t)=1$, such that $\theta{(t+\Delta t)} = 0$ as shown in Eq. \ref{eq: action_update}. This action effectively aligns the agent’s heading directly towards the home location.\\
 The results show a steady increase in $\langle n \rangle$ with increasing $D_r$, continuing even beyond the optimal value $D_r^\ast \sim 12$
 as shown in Fig. \ref{fig:resetting_analysis}(a). From this analysis, we define an effective resetting rate, $\nu = \left\langle \frac{n}{T_{\rm home}} \right\rangle$, which provides a measure of how frequently resettings occur for each $D_r$. For this we computed the quantity $(\frac{n}{T_{\rm home}}$) for each realization and subsequently averaged over all realizations. We observe in Fig. \ref{fig:resetting_analysis}(b) that $\nu$ increases with $D_r$, indicating that higher rotational noise leads to 
 more frequent reset events.
Plotting the mean homing time, $\langle T_{\mathrm{home}} \rangle$, as a function of $\nu$ reveals that it attains a maximum at $\nu^\ast \sim 218$
 as shown in Fig. \ref{fig:resetting_analysis}(c). This value of $\nu^{*}$ corresponds to the optimal $D_r^{*}$, implying their mutual dependence. We also calculated the probability distribution of time interval between two successive resetting $P(\tau, D_r)$, it shows an exponential decay as shown in Fig. \ref{fig:resetting_analysis}(d). The characteristic reset time $\tau^{*}$, obtained from an exponential fit to the probability distribution of time intervals between consecutive resets, is plotted as a function of $D_r$ and found to follow a power-law dependence as shown in Fig. \ref{fig:resetting_analysis}(e). For $D_r \le D_r^{*}$, the $P(\tau, D_r)$ shows a long tail with characteristic time $\tau^{*} \sim \frac{1}{\sqrt{D_r}}$, whereas for $D_r > D_r^{*}$ the $\tau^{*}$ decays as $1/D_r$. Hence for large $D_r$, the characteristic reset time simply follows the rotational noise, whereas for smaller $D_r$ it shows a competition between  rotational noise and stochastic learning. Such two distinct scaling clearly suggests the non-trivial dynamics of the agent for the small and large $D_r$ regimes. In Fig. \ref{fig:resetting_analysis} (e) (inset) we show the two distinct scaling functions  $f(x) \sim e^{-c x}$ ($c$ is a positive constant),  for $D_r > D_r^*$ and $D_r \leq D_r^{*}$, where $x = \tau \sqrt{D_r}$ for $D_r \le D_r^{*}$, whereas $x = \tau D_r$ for $D_r > D_r^{*}$.\\

 To understand the non-monotonic behavior of $\langle T_{\mathrm{home}}\rangle$ and how rotational noise influences the decision-making dynamics, we examine the number of times the two actions given in Eq. \ref{eq: action_update} are selected across different $D_r$, averaged over time and realizations, as shown in Fig. \ref{fig:resetting_analysis}(f). Here, the plotted quantity represents the frequency of action selection rather than the instantaneous value of the action variable itself. The results show a clear crossover in action selection: with increasing $D_r$, the agent increasingly favors action 1, while the preference for action 2 decreases. Physically, larger rotational diffusion leads to stronger angular deviations during action 2, increasing the probability that the system enters the unfavorable state $s(t)=1$ as can be seen from Eq. \ref{eq: state_description}, where the particle tends to move away from the home direction and incurs a larger cost. Through the Q-matrix update as given in Eq. \ref{eq: Q_update}, the RL agent gradually learns to suppress such unfavorable trajectories by increasingly selecting action 1, which acts as a corrective reorientation mechanism. As a result, at larger $D_r$, action 1 progressively dominates over action 2, as shown in Fig. \ref{fig:resetting_analysis}(f). To further clarify this behavior, we have included a separate section ~\ref{3.2} discussing the evolution of the Q-matrix entries corresponding to states 1 and 2 and comparing their behavior across different values of $D_r$. To complement this discussion, Appendix~\ref{Q_entries} presents the complete time evolution of the individual Q-matrix elements, providing additional insight into the learning dynamics throughout the homing process.\\

 The observed crossover reflects adaptive noise compensation: as rotational diffusion strength increases, stochastic angular fluctuations render exploratory reorientation unnecessary, favoring deterministic steering that stabilizes homing, analogous to noise-robust navigation strategies in biological organisms that promotes reliance on deterministic cues (e.g., visual landmarks or chemotactic gradients \cite{buhlmann2011vector, alonso2024learning}) rather than stochastic search, ensuring reliable homing despite noisy dynamics.



 This shift in preference of action selection provides a direct explanation for the observed reduction in the mean homing time $\langle T_{\mathrm{home}} \rangle$ at large $D_r$ as shown in Fig. \ref{fig:mean_thome_vs_Dr}. Since action~1 corresponds to deterministic reorientation towards the home, whereas action~2 introduces stochastic angular updates. The dominance of action~1 at higher $D_r$ therefore suppresses unnecessary stochastic reorientation, leading to more persistent, directed motion and faster convergence to the home region. To further substantiate this mechanism, we analyze the angular dynamics of a representative single realization at different values of $D_r$. The time series of the angular deviation $\theta$ shows that larger fluctuations in $\theta$ are associated with an increased selection of action~2, whereas smaller angular deviations predominantly favor action~1. (See Fig. \ref{fig: delta_theta} in Appendix \ref{effect_delta_theta} for details). Consistently, we find that the standard deviation of the angular deviation, $\sigma_{\theta}$, decreases systematically with increasing $D_r$ after optimal $D_r^{*}$. (See Fig. \ref{fig: standard_delta} in Appendix \ref{standard_single} for details). Despite the increase in the noise strength, the learning dynamics progressively suppress large angular deviations by increasingly favoring action~1, thereby reducing effective orientational fluctuations. This reduction in $\sigma_{\theta}$ explains the diminished use of stochastic reorientation (action~2) and supports the observed acceleration of the homing process at large $D_r$.

In summary, the single-agent results reveal that the homing performance depends non-monotonically on $D_r$. At low noise, trajectories are nearly deterministic with minimal variability, while intermediate $D_r$ leads to the largest fluctuations in $\langle T_{\mathrm{home}} \rangle$ due to strong competition between noise and learning. Beyond a characteristic value $D_r^{*}$, frequent reorientations render the motion diffusive but more statistically uniform, causing both the mean homing time and its spread to decrease. The resetting statistics further support this trend: the mean number of resets and the effective reset rate $\nu$ increase steadily with $D_r$, and the reset-interval distribution yields a characteristic time $\tau^{*}$ that follows a power-law dependence giving two different scaling regimes for $D_r \leq D_r^{*}$ and $D_r > D_r^{*}$. Before moving to the two-agent and multi-agent systems, we first examine the evolution of the Q-matrix elements and the associated action-selection statistics to gain insight into the learning dynamics underlying the homing process.

\subsection{Q-Matrix Evolution and Action Selection}{\label{3.2}}

\begin{figure*}[htbp]
    \centering
    \includegraphics[width=0.8\textwidth]{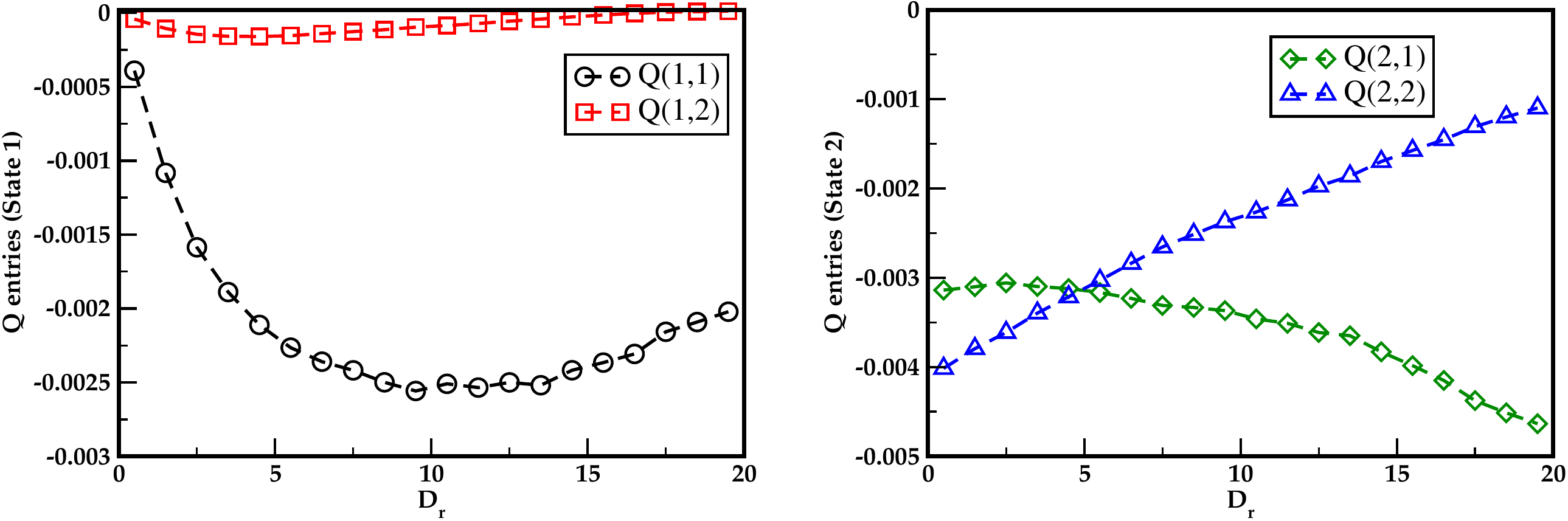}
    \caption{(Color online) Q-matrix elements as a function of rotational diffusion strength $D_r$. The left panel shows the behavior of $Q(1,1)$ and $Q(1,2)$ in the misaligned state ($s=1$), while the right panel shows $Q(2,1)$ and $Q(2,2)$ in the aligned state ($s=2$). The symbols show data points and dashed line is a guide to the eye.}
    \label{fig:qmatrix}
\end{figure*}

To understand the decision-making strategy, we examine the behavior of the Q-matrix elements corresponding to the two possible states and actions, we show the full $2 \times 2$ Q-matrix as a function of the rotational diffusion strength $D_r$, as shown in Fig. \ref{fig:qmatrix}.\\

In each state, the system chooses between two possible actions by comparing their corresponding values. For state 1, comparing ( Q(1,1) ) and ( Q(1,2) ) determines the preferred action, while for state 2, comparing ( Q(2,1) ) and ( Q(2,2) ) determines the preferred action.
In Fig. \ref{fig:qmatrix}, we present the behavior of the Q-matrix elements as a function of $D_r$ in the steady-state regime. The left panel corresponds to state $s=1$, i.e., the misaligned state. In this case, we observe that $Q(1,1)$ remains consistently more negative than $Q(1,2)$ over the entire range of $D_r$. Since lower Q-values correspond to more favorable actions in our cost-based formulation, this implies that action 1 (complete resetting towards the home) is always preferred in the misaligned state. Physically, this is reasonable, as when the particle is misaligned, a strong corrective action is required to redirect it toward the target. At the same time, $Q(1,2)$ remains finite and non-zero, indicating that the alternative action is not entirely suppressed. This confirms that the learned policy is nontrivial and does not reduce to a degenerate form with vanishing entries.\\

The right panel corresponds to state $s=2$, i.e., the aligned state. At low $D_r$, $Q(2,2)$ is more favorable, indicating that action 2 (maintaining or weakly adjusting alignment) is preferred when rotational noise is small. However, as $D_r$ increases, a clear crossover occurs, beyond which $Q(2,1)$ becomes more favorable. This reflects the increasing influence of rotational noise, which makes it advantageous to switch to a stronger corrective action even in the aligned state. Importantly, all Q-matrix elements remain finite across the full range of $D_r$, demonstrating that both actions remain dynamically relevant and that the policy continuously adapts to noise rather than collapsing to a trivial structure.

Overall, the figure highlights that the Q-matrix encodes a nontrivial, noise-dependent decision-making strategy at late times, with state-dependent preferences and smooth transitions between competing actions. Now we study the results for the system with more than one agent.

\subsection{Two-agents Regime}{\label{3.3}}
 \begin{figure}[hbtp]
    \centering
    \includegraphics[width=0.45\textwidth]{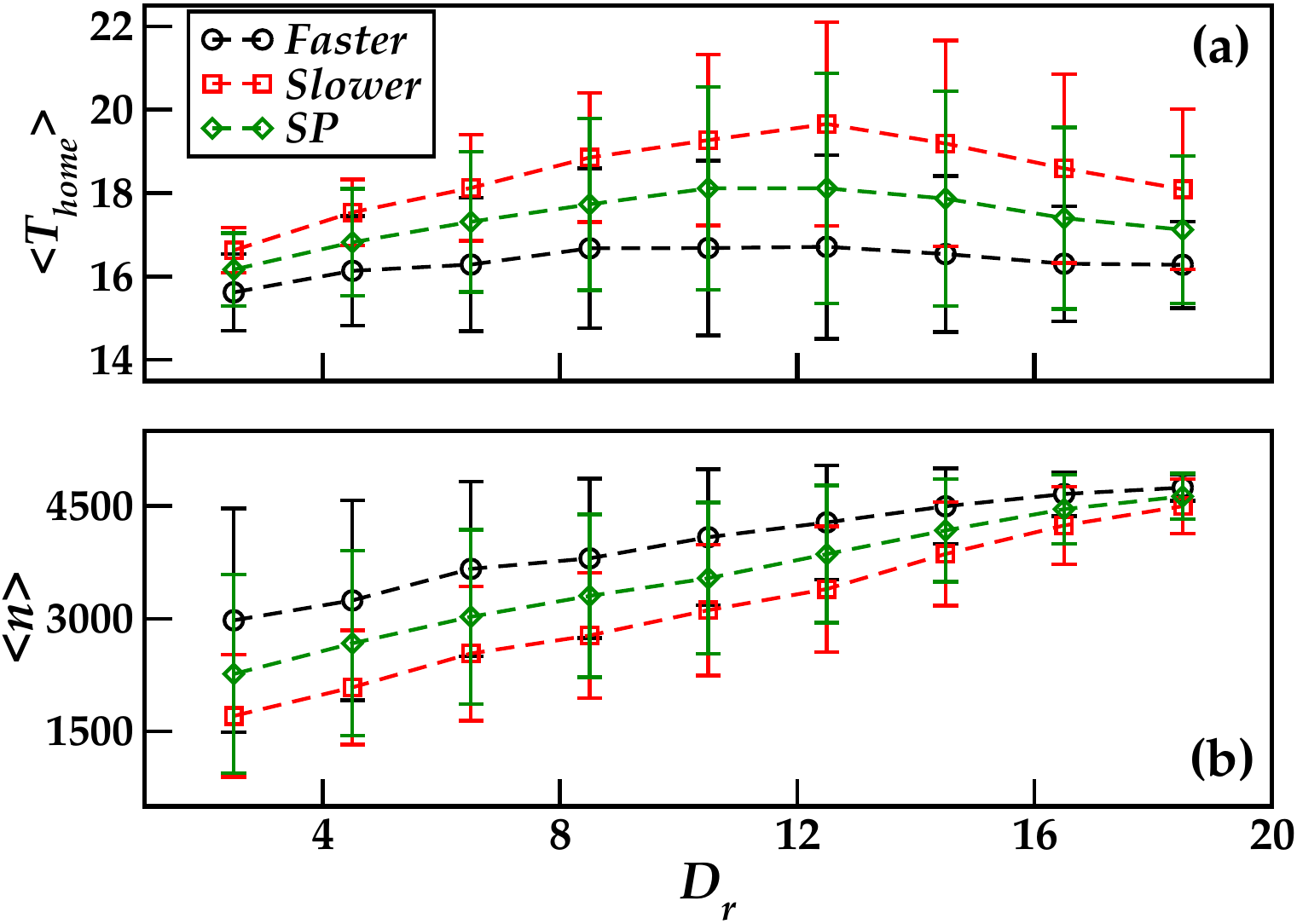}
     \caption{
    (a) (Color online) Plot shows $\langle T_{\mathrm{home}} \rangle$ $vs.$ $D_r$ where faster and slower corresponds to faster and slower agents in two-agents system and SP stands for single particle case. 
    (b) Mean number of resetting events $\langle n \rangle$ plotted with respect to $D_r$ where symbols have same meaning as in (a) panel. Error bars show the standard deviation in respective quantities in each plot.
    }
    \label{fig:two_agents}
\end{figure}

In the two-agents Q-learning system, the homing dynamics exhibit significant variability between the two agents due to their interactions. As shown in Fig. \ref{fig:two_agents}(a), in a group of two agents, one particle consistently reaches the home location faster than the other, even outperforming the single-agent case, while the other particle takes the longer time to arrive. Fig.                \ref{fig:two_agents}(b) illustrates the number of resets for each particle: the faster particle experiences a higher number of resets compared to the single-agent system, whereas the slower particle undergoes fewer resets. Since resets correspond to deterministic reorientations toward the home direction, a higher reset count indicates more frequent corrective alignment, enabling rapid radial progress and thus shorter homing times. In contrast, the slower particle and single agent experience fewer resets, leading to prolonged angular wandering and delayed arrival as shown in Movie 1 (Appendix \ref{movies}).

\begin{figure*}
    \centering
    \includegraphics[width=1.0\linewidth]{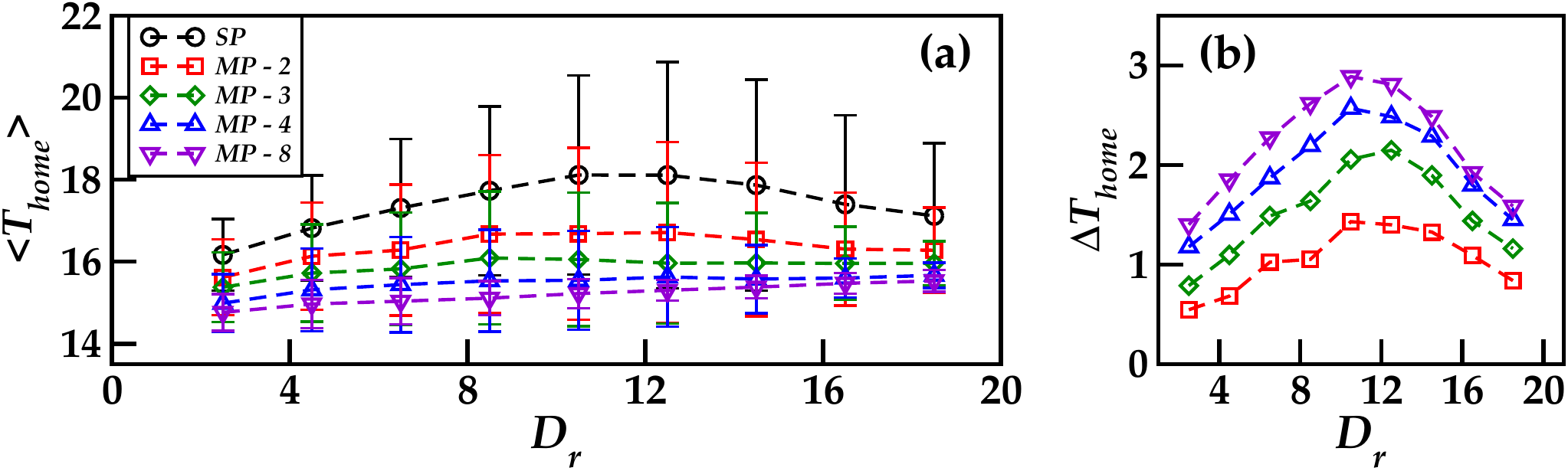}

    \caption{(Color online) (a) $\langle T_{\mathrm{home}} \rangle$ $vs.$ $D_r$ for different population sizes. SP denotes the single-particle case, while MP–$N$ represents the fastest particle in a group of $N$ agents ($N=2,3,4,8$). 
    This plot highlights how the fastest particle benefits from increasing group size, showing a systematic decrease in homing time compared to smaller groups and single-particle dynamics. Error bars show the standard deviation in $\langle T_{\text{home}} \rangle$ in each case. (b) For clearer visualization, the difference in $\langle T_{\mathrm{home}} \rangle$ between the fastest agent in MP–$N$ systems ($N=2,3,4,8$) and the single-particle case (denoted by \textit{$\Delta T_{\mathrm{home}}$}) is shown; symbols correspond to those in panel (a).}
    \label{fig:Thome_multi}
\end{figure*}

This distinction is further quantified by the standard deviation of the angular deviation, $\sigma_\theta$, which decreases systematically with increasing $D_r$ beyond the optimal noise strength $D_r^{*}$. (See Fig. \ref{fig:std_theta_comparison} (a) in Appendix \ref{standard_multi} for details). Notably, the suppression of orientational fluctuations is most pronounced for the faster particle in the two-agents system, whose $\sigma_\theta$ is significantly smaller than that of both the slower particle and the single-agent case.

In conclusion, the emergence of a faster particle in the two-agent system can be traced to the combined effects of reset statistics and angular stability. The faster particle experiences more frequent resets, leading to deterministic reorientations toward the home direction, which suppress angular fluctuations in $\theta(t)$ and yield a significantly smaller standard deviation $\sigma_\theta$ compared to both the slower particle and the single-agent case. The suppression in orientational fluctuations  minimizes angular wandering and sustains persistent radial motion, resulting in faster homing.

\subsection{Multi-agents Regime}{\label{3.4}}

\begin{figure}[h]
    \centering
    \includegraphics[width=1.0\linewidth]{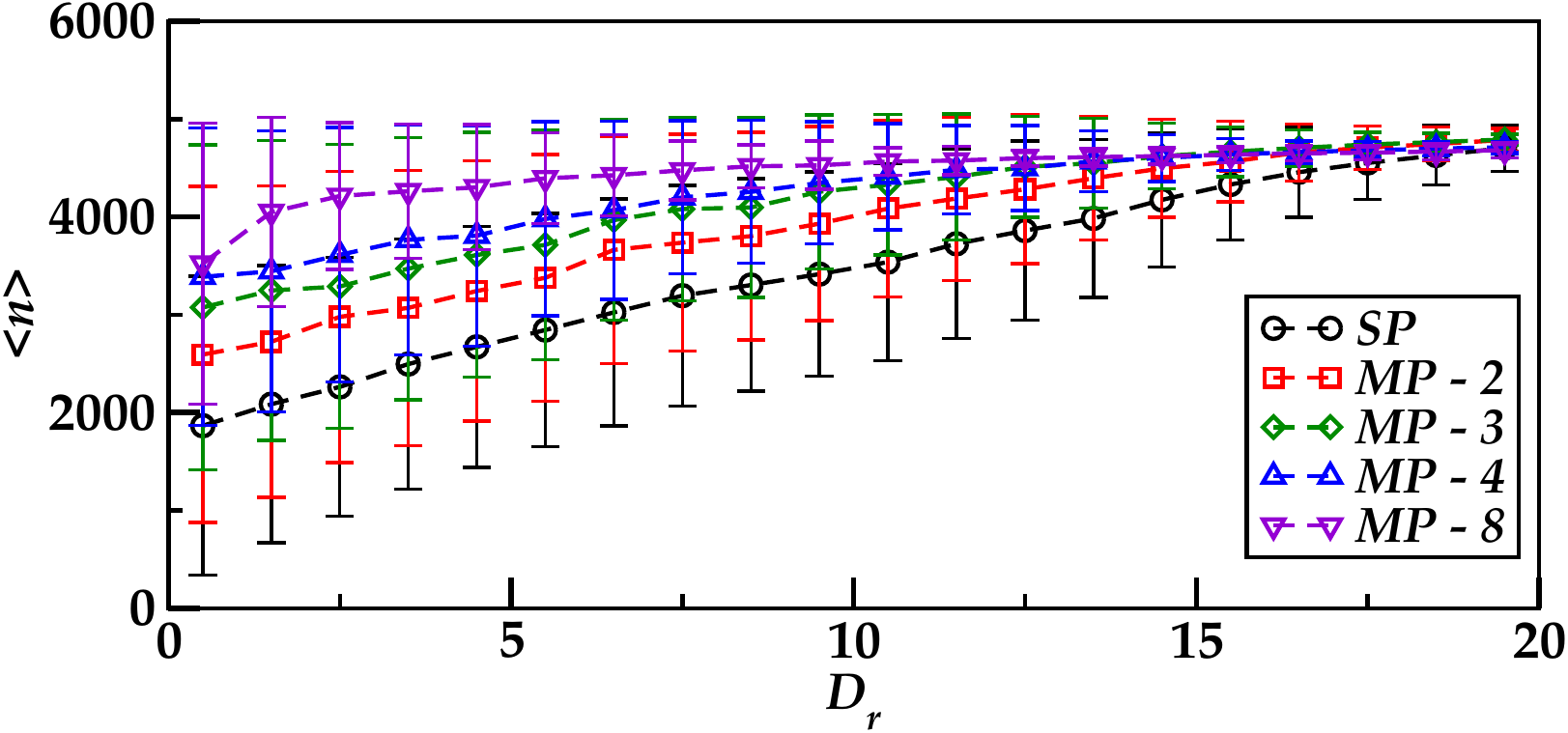}
    \caption{ (Color online) Mean number of resetting events $\langle n\rangle$ plotted with respect to $D_r$. Error bars show the standard deviation in respective quantities in each plot.} 
    \label{fig:multi_resets}
    \end{figure}

For the multi-agents Q-learning system, we observe in Fig. \ref{fig:Thome_multi}(a) that as the number of particles in the group increases, the homing time for the fastest particle decreases. This indicates that the fastest particle becomes increasingly faster in larger groups. For clearer visualization of this effect, the difference in the homing time $\Delta  T_{\text{home}} $ between the single-particle and multi-particle cases is plotted as a function of $D_r$ in Fig. \ref{fig:Thome_multi}(b). The reduction in the homing time of the fastest particle cannot be attributed solely to the presence of a larger number of particles, which by itself would increase the likelihood of observing an exceptionally fast trajectory by chance. In the present model, particles interact through soft repulsive forces, which modify their trajectories and orientation dynamics. These interaction-induced changes modify the learning process and are accompanied by an increase in the mean number of resetting events with population size, as shown in Fig.~\ref{fig:multi_resets}. The more frequent resetting enhances the ability of particles to reorient toward the home region, thereby promoting faster homing.

Notably, the reduction in angular fluctuations in $\theta(t)$ quantified by standard deviation $\sigma_\theta$, is strongest for the fastest particle when the group size is large (See Fig. \ref{fig:std_theta_comparison}(b) in Appendix \ref{standard_multi} for details). This reduction arises from short-range repulsive interactions between agents. These interactions promote frequent resetting, which suppresses large orientational deviations and repeatedly realigns the fastest agent towards homing direction. In contrast, when the group size is reduced, resetting events become less frequent as can be seen in Fig. \ref{fig:multi_resets}, and orientational fluctuations become more pronounced, resulting in diminished directional persistence and reduced performance.

\subsection{RL Vs ABP (with resetting)}: Comparison of Homing Efficiency{\label{3.5}}
To provide a baseline for comparing homing performance between Reinforcement Learning (RL)--based agents and stochastic dynamics, we introduce an Active Brownian Particle (ABP) model with resetting. The goal is to compute the mean homing time $\langle T_{\mathrm{home}} \rangle$ of a non-learning particle so that the RL performance can be quantitatively compared against it. In this model, the particle’s movement arises solely from self-propulsion, stochastic angular diffusion, and a predefined state-dependent resetting probability given as :

\begin{equation}
p(t) = 
\begin{cases}
0, & |\theta| \le \phi, \\[8pt]
\dfrac{|\theta| - \phi}{\pi - \phi}, & \phi < |\theta| < \pi, \\[10pt]
1, & |\theta| \ge \pi.

\end{cases}
\label{eq:probability}
\end{equation}

The functional form of probability is motivated by the need to provide a smooth transition between aligned and misaligned configurations. When the particle is well aligned ($|\theta| \le \phi$), no resetting is required, and hence $p(t)=0$. As the angular deviation increases beyond the threshold $\phi$, the probability of resetting increases linearly, reflecting a growing need for corrective action. For large deviations ($|\theta| \ge \pi$), resetting becomes certain ($p(t)=1$), as the particle is maximally misaligned.

Thus, Eq. \ref{eq:probability} provides a simple yet physically meaningful rule that links the degree of misalignment to the likelihood of taking a corrective action. Thus, the equation captures the balance between persistent exploration and corrective reorientation in a simple probabilistic manner.

At every time step, the particle selects one of the two possible orientation-update actions based on the above probability $p(t)$.
	
\begin{equation}
\theta{(t+\Delta t)} =
\begin{cases}
0, & \text{with probability } p(t), \\[8pt]
\theta(t) + \sqrt{2D_r \Delta t}\,\eta,&\text{with probability }1 - p(t).
\end{cases}
\end{equation}

where $\eta(\mathbf{r},t)$ is a Gaussian white noise with zero mean,
$\langle \eta(\mathbf{r},t) \rangle = 0$, and correlations given by:
\begin{equation}
\langle \eta(\mathbf{r},t)\,\eta(\mathbf{r}',t') \rangle
= \delta(\mathbf{r}-\mathbf{r}')\,\delta(t-t').
\end{equation}

All other update rules for position and orientation follow the same definitions as in the RL model however, the ABP performs these updates without any learning mechanism—there is no policy optimization, no cost-based adaptation, and no Q-matrix update, and the homing time is computed using the same criterion as in RL. The meanings of all symbols remain identical to those introduced in the RL formulation.
By comparing the mean homing time \(\langle T_{\mathrm{home}} \rangle\) for ABP with resetting and RL across a range of rotational diffusion strengths \(D_r\), the RL trajectories are found to be shorter and less noisy, leading to consistently 
\[
\langle T_{\mathrm{home}} \rangle_{\mathrm{RL}} < \langle T_{\mathrm{home}} \rangle_{\mathrm{ABP}}.
\]
For better visualization of comparison of particle trajectory for RL-agent and ABP with resetting, we provide movies of a moving RL-agent and ABP for $D_r = 2$, $12$  and  $15$. The link for the movie can be accessed through Appendix \ref{movies} (Movies: 2, 3 and 4). We clearly see much more wandering of ABP particle in comparison to RL agent. 
\begin{figure}[h]
    \centering
    \includegraphics[width=1.0\linewidth]{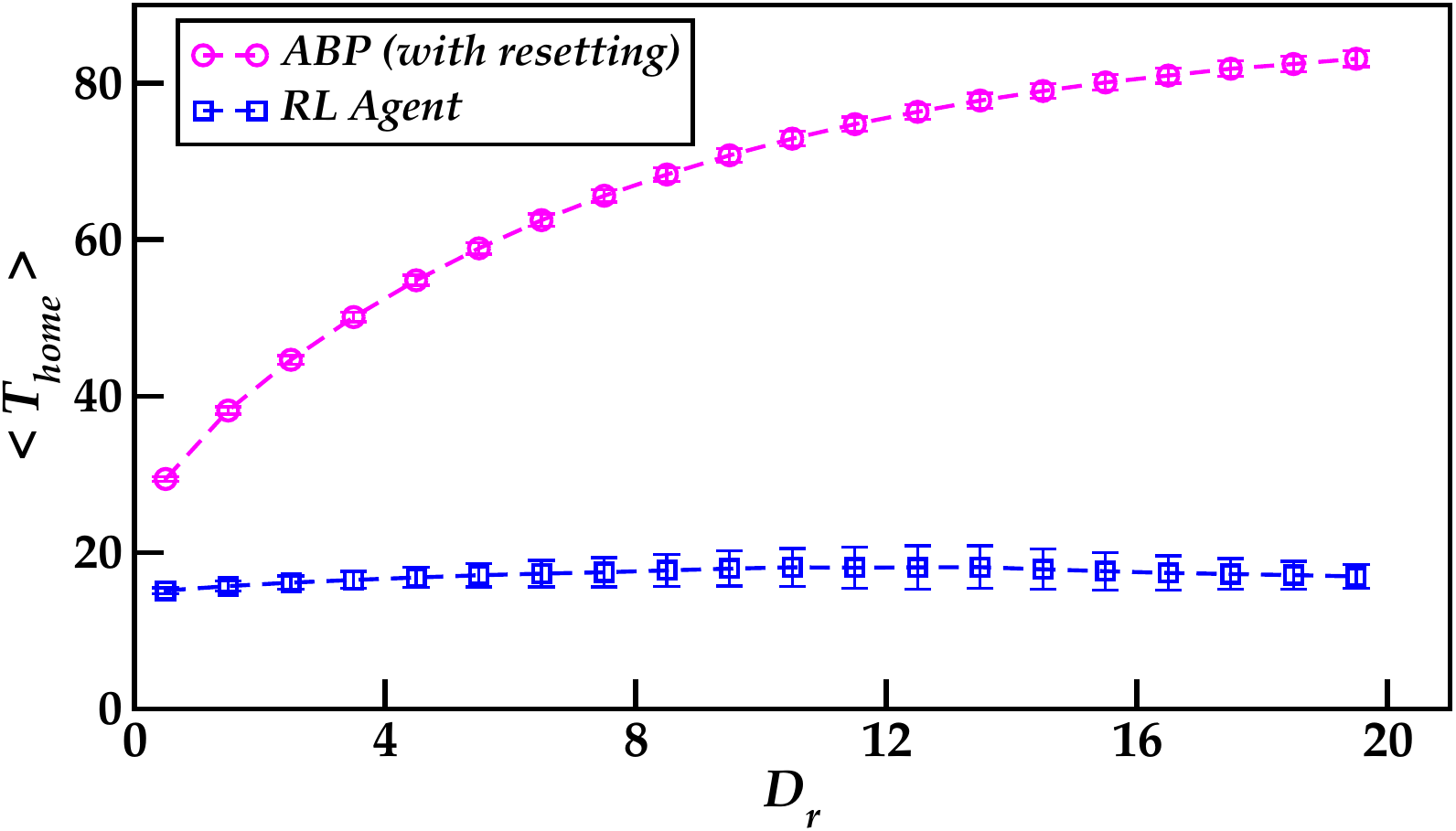}
    \caption{(Color online) \(\langle T_{\mathrm{home}} \rangle\) $vs.$ $D_r$ along with error bars for an ABP with resetting and a RL agent. The RL trajectories exhibit consistently shorter homing times than ABP across all noise levels, highlighting the improvement due to learning.}
    \label{fig:abp}
    \end{figure}

\subsection{Pure ABP (without resetting)}\label{3.6}

For completeness, we have also analyzed the behavior of a pure ABP without resetting. In this case, the particle undergoes persistent motion with rotational diffusion but lacks any mechanism for directed correction towards the home.
The particle position is evolved using the same kinematic rule as used in the RL and ABP with resetting cases, while the orientation evolves through rotational diffusion according to:
\begin{equation}
\
\theta(t+\Delta t)=\theta(t)+\sqrt{2D_r\Delta t}\,\eta,
\
\end{equation}

The meanings of all symbols remain identical to those introduced previously and the homing
time is computed using the same criterion as in RL and ABP with resetting cases. Since no resetting or corrective alignment mechanism is present , the dynamics are governed entirely by the interplay between persistent self-propulsion and stochastic rotational diffusion and thus lacks an adaptive mechanism to suppress unfavorable trajectories or correct large angular deviations.\\

For comparison, we have also attached a movie showing the trajectories for the reinforcement learning (RL) agent, ABP with resetting, and pure ABP (without resetting) at a particular value of $D_r$ = 2. This highlights the importance of resetting to ensure reliable convergence to the home. The link for the movie can be accessed through Appendix \ref{movies} (Movie: 5). We clearly see, the RL agent reaches the target roughly three times faster than the ABP with resetting and about eight times faster than the pure ABP (without resetting), highlighting the enhanced efficiency of the learned adaptive strategy. In contrast, the pure ABP lacks any corrective feedback mechanism and therefore spends long times wandering due to stochastic rotational diffusion, leading to significantly slower and less reliable convergence toward the target.    

\section{Conclusions}{\label{4}}

In this work, we employed a reinforcement learning--based approach to investigate homing dynamics in active agents, progressing systematically from a single-agent system to interacting two-agent and multi-agent configurations with short-range, spring-like repulsive interactions. The primary objective is to minimize the homing time while accounting for stochastic reorientations. For a single agent, we observe that the mean homing time $\langle T_{\mathrm{home}} \rangle$ exhibits a clear non-monotonic dependence on the rotational diffusion strength $D_r$, revealing an optimal noise level $D_r^\ast$ that balances stochastic exploration and learned directional control. At low noise, trajectories are nearly deterministic, whereas intermediate noise leads to enhanced fluctuations due to competition between learning and randomness. Beyond $D_r^\ast$, larger reorientations leads to larger cost and results in more frequent resetting and hence reduces both the mean homing time and its variability.

The resetting statistics further support this picture: both the mean number of resetting events and the effective resetting rate increase monotonically with $D_r$, while the distribution of reset intervals yields a characteristic time scale $\tau^\ast$ that follows a power-law dependence on $D_r$, with two distinct scaling regimes for $D_r \leq D_r^{*}$ and $D_r > D_r^{*}$.

Furthermore, we have compared the homing time obtained from the RL agent with that of an Active Brownian Particle (ABP) with resetting and a pure ABP (without resetting) under identical conditions. The RL-based agent consistently achieves shorter homing times than both cases, while the pure ABP typically continues wandering near the target without reliable localization. Extending the model to interacting agents, we find that asymmetry naturally emerges in two-agents system, with one agent consistently achieving faster homing. In larger populations, repulsive interactions maintain spatial separation while collective effects increasingly favor the fastest agent as group size increases. Overall, our results demonstrate that reinforcement learning effectively captures the interplay between stochasticity, resetting, and inter-agent interactions,
providing a scalable framework to study efficient homing and navigation in both solitary and collective systems.

In the present work, we have defined the agent’s state as the difference between the homing direction and its instantaneous orientation, which provides a simple yet effective representation of the agent’s alignment with the target.
Alternative formulations, such as polar states $(r, \theta)$ as well as velocity-based states can provide a more dynamic sense of progress. However, both these approaches increase the dimensionality of the state space and slow down convergence.
Hence, our chosen state definition strikes a practical balance between information content and computational efficiency, allowing the agent to learn effective homing strategies with minimal complexity. \\

We have also examined a simple alignment interaction with a vision cone of $90^\circ$. Although alignment improved collective motion in some regimes, it did not help the agents locate the target earlier over the full range of $D_r$, particularly at larger noise strengths. This suggests that global alignment alone is not sufficient for efficient target finding, and therefore instead of aligning with all agents in the group one can follow the fastest member of the group. Therefore, in future extensions of this work, the framework can be advanced by introducing a leader–follower learning mechanism. Such an approach would enable the emergence of collective intelligence and coordinated navigation, allowing slower agents to improve their performance by leveraging information extracted from the most efficient member of the group.

\section*{\textbf{Acknowledgment}}
 R S, P J, A K and S M
 thank PARAM Shivay for the computational facility under the National Supercomputing Mission,
Government of India, at the Indian Institute of Technology (BHU) Varanasi and also IIT (BHU) Varanasi
computational facility. S.M. thanks
DST, SERB (INDIA), Project No. CRG/2021/006945 and MTR/2021/000438 for partial financial support.\\

\section*{\textbf{Conflict of interest declaration}}
The authors declare that there are no conflicts of interest.
\section*{\textbf{Data availability statement}}
All data that support the findings of this study are included within the article (and any supplementary files).


\onecolumngrid

\appendix
\section{Dynamics of two and multi-agent systems}    \label{Dynamics:two_multi}
\subsection{Two-agents regime}

For the two-agents system, the particles are initially placed at two nearby locations, slightly away from the boundary and at approximately the same radial distance from the home position. The initial distance of each agent from the home is chosen in the range $r_0 \sim 27$. Each particle is assigned a random initial orientation to ensure stochasticity in the initial conditions.
 All components of the RL scheme - state definitions, $\varepsilon$-greedy action selection, and cost-based Q-table updates - are identical to those used in the single-agent model. The only modification in the two-agents setup is the introduction of finite particle size and short-range repulsive interactions.

In the numerical implementation, the particle positions are updated according to

\begin{equation}
\mathbf{r}_i(t+\Delta t)
= \mathbf{r}_i(t)
+ v_0\,\Delta t\,\hat{\mathbf{n}}_i(t+\Delta t)
+ \Delta t \sum_j \mathbf{F}_{ij},
\label{eq: position}
\tag{A1}
\end{equation}
where $\mathbf{r}_i(t) = (x_i(t),\, y_i(t))$ is the position of particle $i$, and the unit orientation vector at the updated time is $\hat{\mathbf{n}}_i(t+\Delta t) = (\cos\theta_i(t+\Delta t),\, \sin\theta_i(t+\Delta t))$.
The values of the self-propulsion speed $v_0$ and the time step $\Delta t$ are identical to those used in the single-agent case. The resulting interaction force on particle $i$ is $\mathbf{F}_{ij} = -\nabla U(r_{ij})$ where $U(r_{ij})$ is harmonic potential defined as:
\begin{equation}
U(r_{ij}) = \frac{\kappa}{2}\,(r_{ij}-\sigma_{ij})^2\,\Theta\!\left(1 - \frac{r_{ij}}{\sigma_{ij}}\right),
\label{eq: potential}
\tag{A2}
\end{equation}
where $\Theta(x)=1$ for $x \ge 0$ and $0$ otherwise,  
$r_{ij} = |\,\mathbf{r}_i - \mathbf{r}_j\,|
$ is the inter-particle separation, and $\kappa$ is the force constant having value 70. Each particle is assigned a radius $R_i = 0.35$, and repulsion acts only when two particles approach closer than their contact separation $\sigma_{ij} = R_i + R_j$.
Repulsive interactions are disabled for any particle that has already reached the home location.

\subsection{Multi-agents regime}

For systems with three, four, and eight particles, the initial positions of the agents are chosen such that no particle begins significantly closer to or farther from the home location than the others, while maintaining sufficient spatial separation to avoid immediate overlap. The initial distance of each agent from the home is chosen in the range $r_0 \sim 27$. In all cases, each particle is assigned a random initial orientation to ensure stochasticity in the starting configuration.

All RL components — including the state definitions, \(\varepsilon\)-greedy action selection, cost-based Q-value updates, the position update rule \ref{eq: position} and treatment of finite-sized particles with short-range repulsive interactions \ref{eq: potential} — are identical to those used in the two-agents system. No additional learning parameters are introduced in the multi-agents case.

\section{Parameter Optimization: Effect of Exploration Probability ($\varepsilon$) and Learning Rate ($\alpha$) on Homing Dynamics}
\label{Parameters}

\begin{figure*}[htbp]
    \centering
 \includegraphics[width=1.0\linewidth]{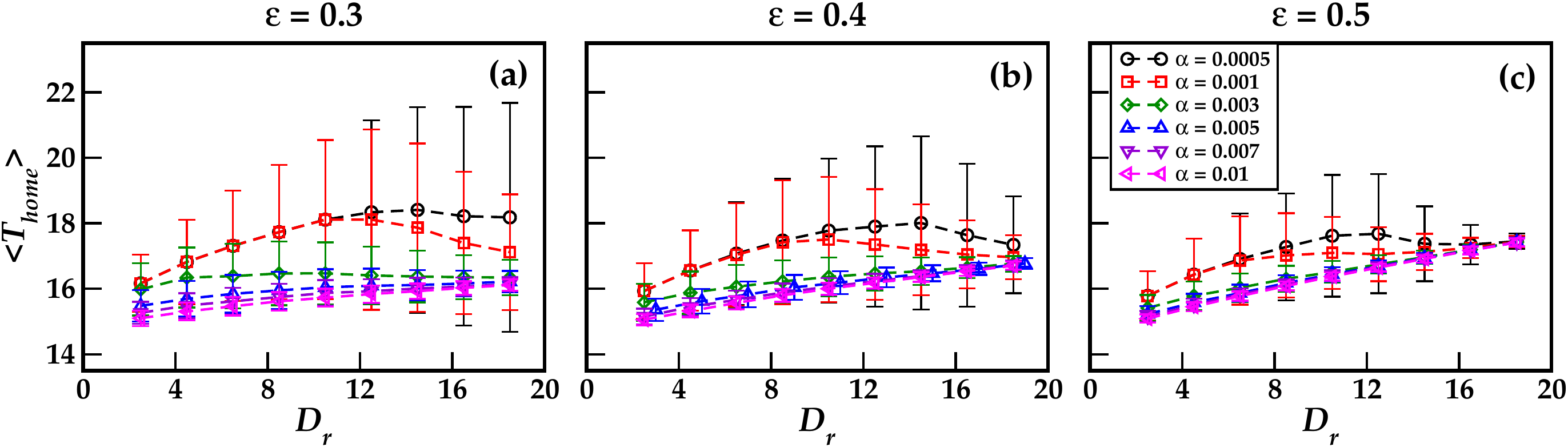}
    \caption{(Color online) The plots (a-c) show the variation of the mean homing time $\langle T_{\mathrm{home}} \rangle$ against rotational diffusion strength $D_r$. The three panels correspond to different exploration probabilities $\varepsilon$ = 0.3, 0.4, and 0.5, while each curve within a panel represents a distinct learning rate $\alpha$ ranging from $0.0005$ to $0.01$. The symbols show the data points and  the dashed line is a guide to the eye while error bars indicate the corresponding standard deviations. }

    \label{fig:Thome_vs_Dr_epsilon_alpha}
\end{figure*}

In this section, we analyze the dependence of the mean homing time $\langle T_{\mathrm{home}} \rangle$ on the rotational diffusion strength $D_r$ for different combinations of exploration and learning parameters. Specifically,   $\langle T_{\mathrm{home}} \rangle$ is plotted as a function of $D_r$ for three values of exploration probability $\varepsilon = 0.3,~0.4,$ and $0.5$, while the learning rate $\alpha$ is varied in the range $\alpha = 0.0005,~0.001,~0.003,~0.005,~0.007,$ and $0.01$ as shown in Fig. \ref{fig:Thome_vs_Dr_epsilon_alpha} (a-c). This systematic exploration allows us to identify the most suitable combination of $(\varepsilon, \alpha)$ parameters that produce optimal homing performance for subsequent analysis.

\section{Time Series of Q-Matrix Elements}
\label{Q_entries}

In this section, we examined the temporal evolution of the Q-matrix during the homing process. We find that different state-action pairs evolve differently depending on their contribution to successful navigation. In particular, actions that are consistently unfavorable quickly acquire lower Q-values, whereas the values associated with actions relevant for successful homing continue to evolve as the agent accumulates experience.

\begin{figure*}[htbp]
\centering
\includegraphics[width=0.80\textwidth]{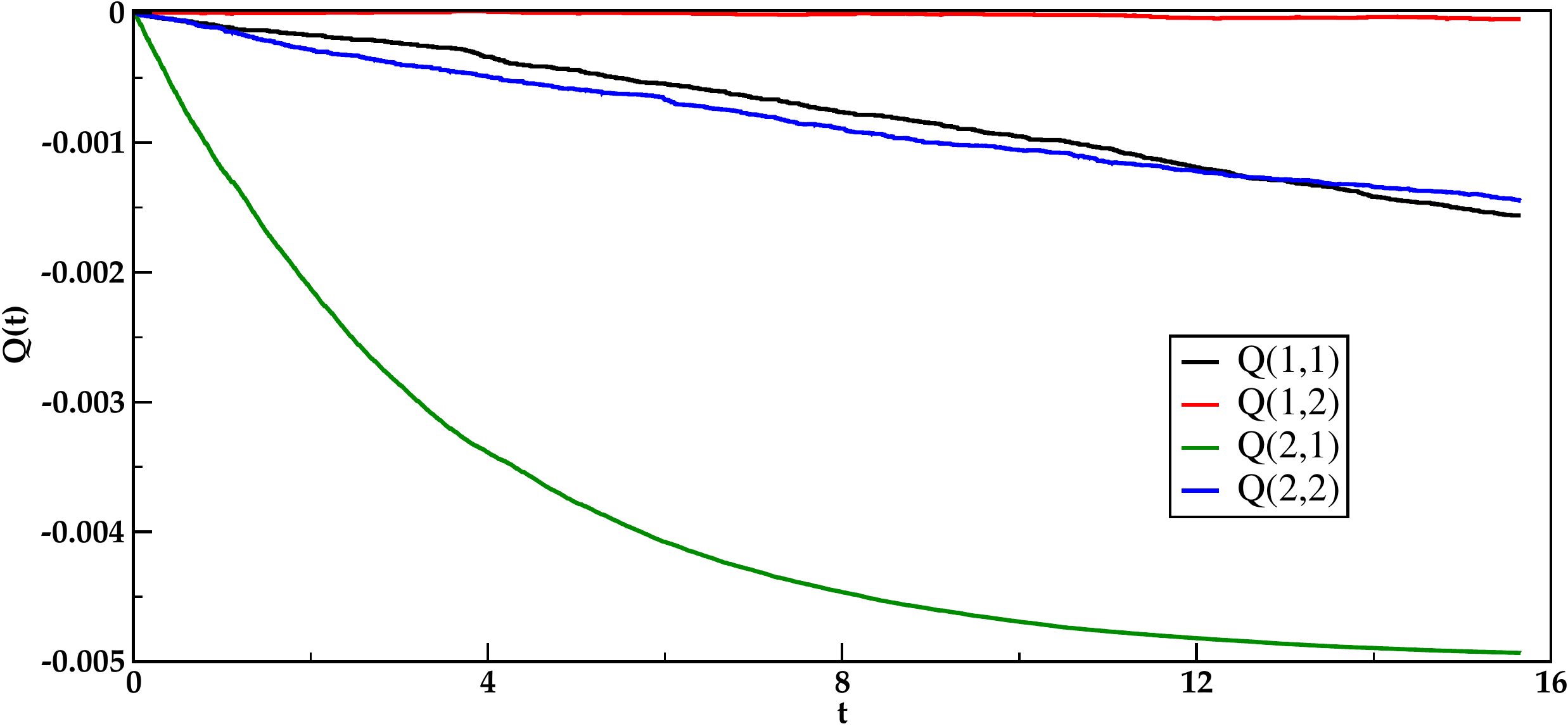}
\caption{ Time evolution of the Q-matrix elements for the four state-action pairs during a representative homing process for a single realization at optimal $D_r^{*}=12$. The Q-values develop distinct trends as the agent acquires experience. Actions that are unfavorable for homing rapidly attain lower Q-values, whereas actions associated with successful navigation continue to be updated throughout the trajectory. The evolution of the Q-matrix illustrates how the learned preference for different actions emerges during the homing process.}
\label{fig:time_series_Q_matrix}
\end{figure*}

Fig.\ref{fig:time_series_Q_matrix} shows the evolution of the four Q-matrix elements during a representative homing process for a single realization at optimal $D_r^{*}=12$. Initially, all Q-values are nearly identical since the agent has not yet accumulated sufficient experience regarding the consequences of different actions. As learning proceeds, the Q-values separate and develop distinct trends depending on the state-action pair. The element Q(2,1), corresponding to resetting in the aligned state, rapidly decreases, indicating that this action is consistently associated with a higher cost. In contrast, Q(1,2), corresponding to maintaining the current orientation in the misaligned state, remains close to zero. The elements Q(1,1) and Q(2,2), corresponding respectively to resetting in the misaligned state and maintaining orientation with some stochasticity in the aligned state, continue to evolve throughout the trajectory because these actions play a more significant role in determining the success of the homing process. Their continued evolution indicates that these action values are still being updated as the agent explores the environment and moves toward the target. The time evolution therefore provides insight into how the agent progressively distinguishes between favorable and unfavorable actions during learning.

\section{Averaging and Error Bar Behavior}
\label{Error_bar}
\begin{figure}[h!]
    \centering
    \includegraphics[width=1.0\textwidth]{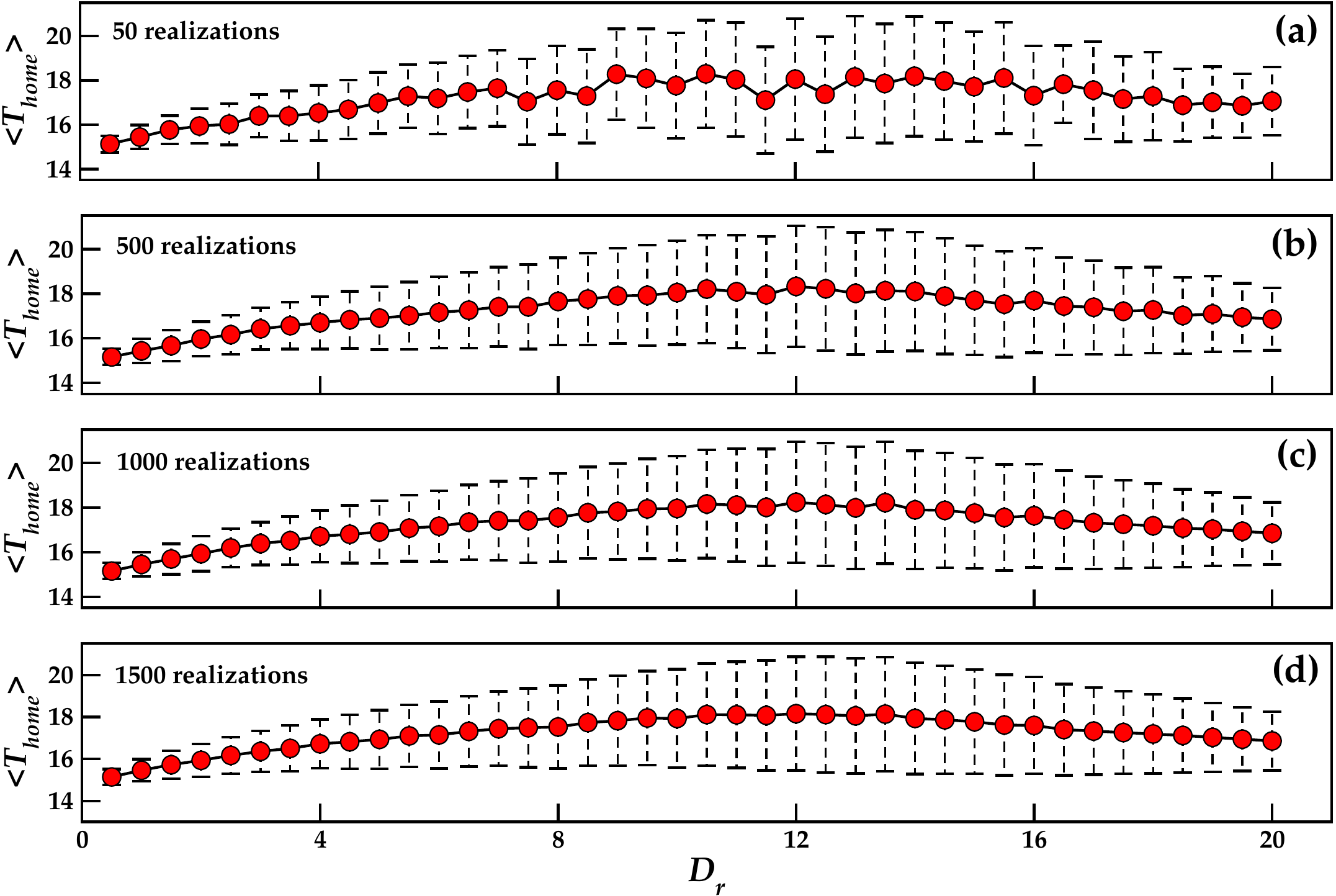} 
    \caption{(Color online) Four panels (a-d) show $\langle T_{\mathrm{home}} \rangle$ $vs.$ $D_r$ for $\varepsilon = 0.3$ and $\alpha = 0.001$ corresponding to different realizations (50, 500, 1000, and 1500). The error bars represent the standard deviation in $\langle T_{\mathrm{home}} \rangle$ for each $D_r$ value within the respective realization.}

    \label{fig:Thome_errorbars_ensembles}
\end{figure}

To examine whether the length of the error bars indicating the corresponding standard deviations of the homing-time distribution obtained from independent trajectories is an intrinsic property of the system or can be reduced by averaging over different realizations, we plot $\langle T_{\mathrm{home}} \rangle$ 
$vs.$ $D_r$ for $\varepsilon = 0.3$ and $\alpha = 0.001$ across different independent realizations: 50, 500, 1000, and 1500 as shown in Fig. \ref{fig:Thome_errorbars_ensembles} (a-d). The results indicate that increasing realizations does not reduce the error bars; rather the trajectories become more synchronized, suggesting that the variability captured by the error bars reflects inherent system fluctuations.

\begin{figure}[hbtp]
\centering
\includegraphics[width=0.60\textwidth]{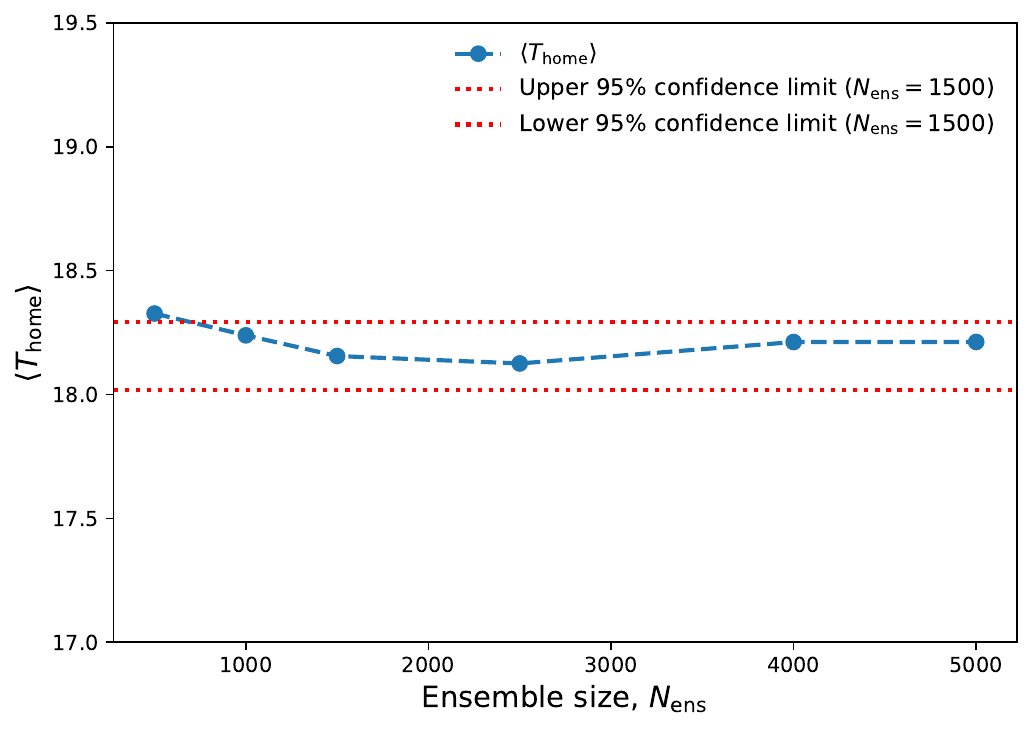}
\caption{
Convergence analysis of the $\langle T_{\rm home} \rangle$ at $D_r^{*}=12$, corresponding to the parameter value where the $\langle T_{\rm home} \rangle$ is largest. The mean homing time is computed using $N_{ens}=500$, $1000$, $1500$, $2500$, $4000$, and $5000$ independent realizations. Although small variations are observed with increasing number of realizations, but the $\langle T_{\rm home} \rangle$ approaches a stable value, indicating statistical convergence. The red dotted horizontal lines indicate the estimated $95\%$ confidence interval, $\langle T_{home} \rangle \pm 1.96\,\sigma/\sqrt{N_{\rm ens}}$, obtained from the $N_{\rm ens}=1500$ realization dataset. For $D_r^{*}=12$, corresponding to the largest mean homing time considered in this work, the mean homing times computed using $N_{\rm ens}=2500$, $4000$, and $5000$ realizations all lie within this confidence interval, indicating that the mean homing time is statistically converged beyond $N_{\rm ens}=1500$.}
\label{fig:convergence}
\end{figure}

To further address the issue of statistical uncertainty associated with ensemble averaging, we performed an additional convergence analysis at $D_r^{*}=12$, corresponding to the parameter value for which the mean homing time is largest and where statistical fluctuations are expected to be most pronounced. We additionally computed the mean homing time using $N_{\rm ens}=2500$, $4000$, and $5000$, independent realizations. While the mean homing time exhibits small variations with increasing number of realizations, the estimates progressively approach a stable value.
For the $N_{\rm ens}=1500$ dataset, we obtained \[ \langle T_{\rm home}\rangle = 18.155. \] The $95\%$ confidence interval of the mean was computed as \[ \langle T_{\rm home}\rangle \pm 1.96\,\frac{\sigma}{\sqrt{N_{\rm ens}}}, \] where $\sigma$ is the standard deviation of the homing-time distribution. Using $N_{\rm ens}=1500$ and the measured value of $\sigma = 2.704$, we obtain \[ 1.96\,\frac{\sigma}{\sqrt{1500}} = 0.137. \] Therefore, the $95\%$ confidence interval is \[ 18.155 \pm 0.137, \] yielding \[ [18.018,\;18.292]. \]
We find that the mean homing times obtained using $N_{\rm ens}=2500$, $4000$, and $5000$ realizations all lie within these confidence bounds. This demonstrates that the estimate obtained from $N_{\rm ens}=1500$ realizations is already statistically reliable and that increasing the number of realizations does not lead to a significant change in the mean homing time. Furthermore, the overall behavior of the results, including the observed trends and physical conclusions, remains qualitatively unchanged upon increasing the number of realizations. Therefore, $N_{\rm ens}=1500$ realizations are sufficient to obtain a statistically converged estimate of the mean homing time.\\
\clearpage
\section{Effect of rotational noise on angular deviation and action selection}
\label{effect_delta_theta}

In order to examine the effect of rotational noise on angular dynamics and action selection, we analyze the time evolution of the angular deviation $\theta(t)$ and the corresponding action-selection behavior for different values of the rotational diffusion strength $D_r$. For a single realization, the left panel illustrates the temporal evolution of $\theta(t)$ for $D_r$ values chosen above and below the optimal $D_r$, while the right panel shows the associated action-selection dynamics, indicating how many times action~1 and action~2 are chosen for the same realization and identical $D_r$ values as in the left panel. Consistent with the observed trends, larger angular fluctuations are correlated with an increased preference for action~2, whereas reduced angular fluctuations are associated with a dominant selection of action~1. This behavior highlights the direct link between suppression in orientational fluctuations and the reinforcement-learning decision-making process.

\begin{figure}[htbp]
    \centering
    \includegraphics[width=1.0\linewidth]{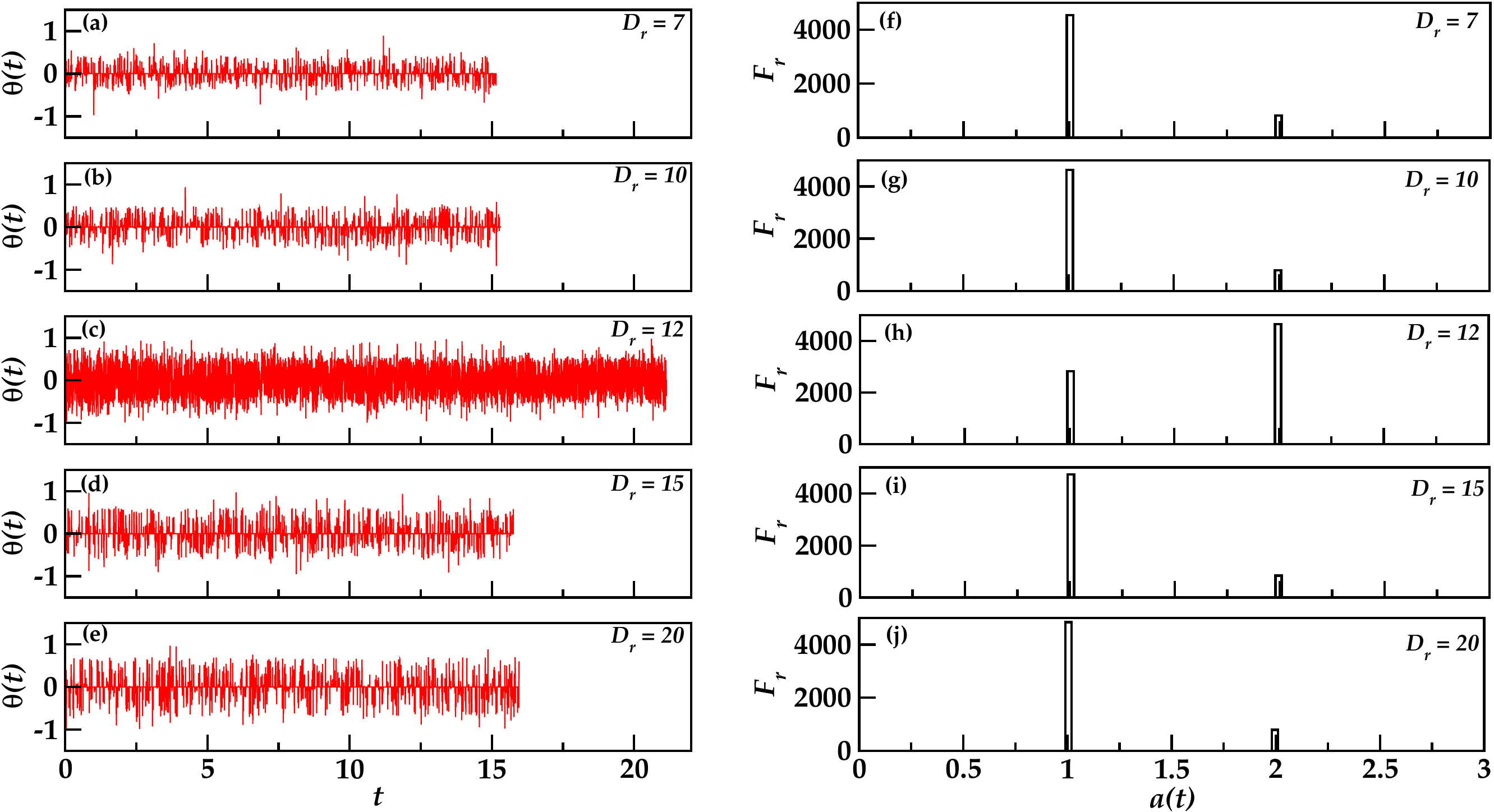}
    \caption{(Color online) The left panel (a-d) shows the time evolution of the angular deviation $\theta(t)$ from the homing direction for different values of the rotational diffusion strength $D_r$, chosen above and below the optimal $D_r$, for a single realization. The right panel (a-d) shows the corresponding action-selection frequency (denoted as \textit{$F_r$}) for the same realization and identical $D_r$ values as in the left panel.}
    \label{fig: delta_theta}
 
\end{figure}

 \section{Standard Deviation $\sigma_{\theta}$ as a Function of $D_r$ : Single Agent}
 \label{standard_single}

To understand the decrease in the mean homing time $\langle T_{\mathrm{home}} \rangle$ beyond optimal $D_r^{*}$, we analyze the standard deviation of the angular deviation $\theta(t)$ as a function of $D_r$. The analysis is restricted to regions beyond a radial distance of $2.5$ units within a disc of radius $R_0 = 35$, where the home radius is fixed at $2$, in order to eliminate the trivial near-home behavior. The calculation is performed over $2000$ independent realizations with a time step $\Delta t = 0.05$. The results show that the standard deviation decreases with increasing $D_r$ beyond the optimal value $D_r^{*}\sim 12$, indicating that frequent resetting lead to a progressively narrower angular distribution $\theta(t)$, corresponding to more localized angular dynamics.

\begin{figure}[htbp]
    \centering
    \includegraphics[width=0.7\linewidth]{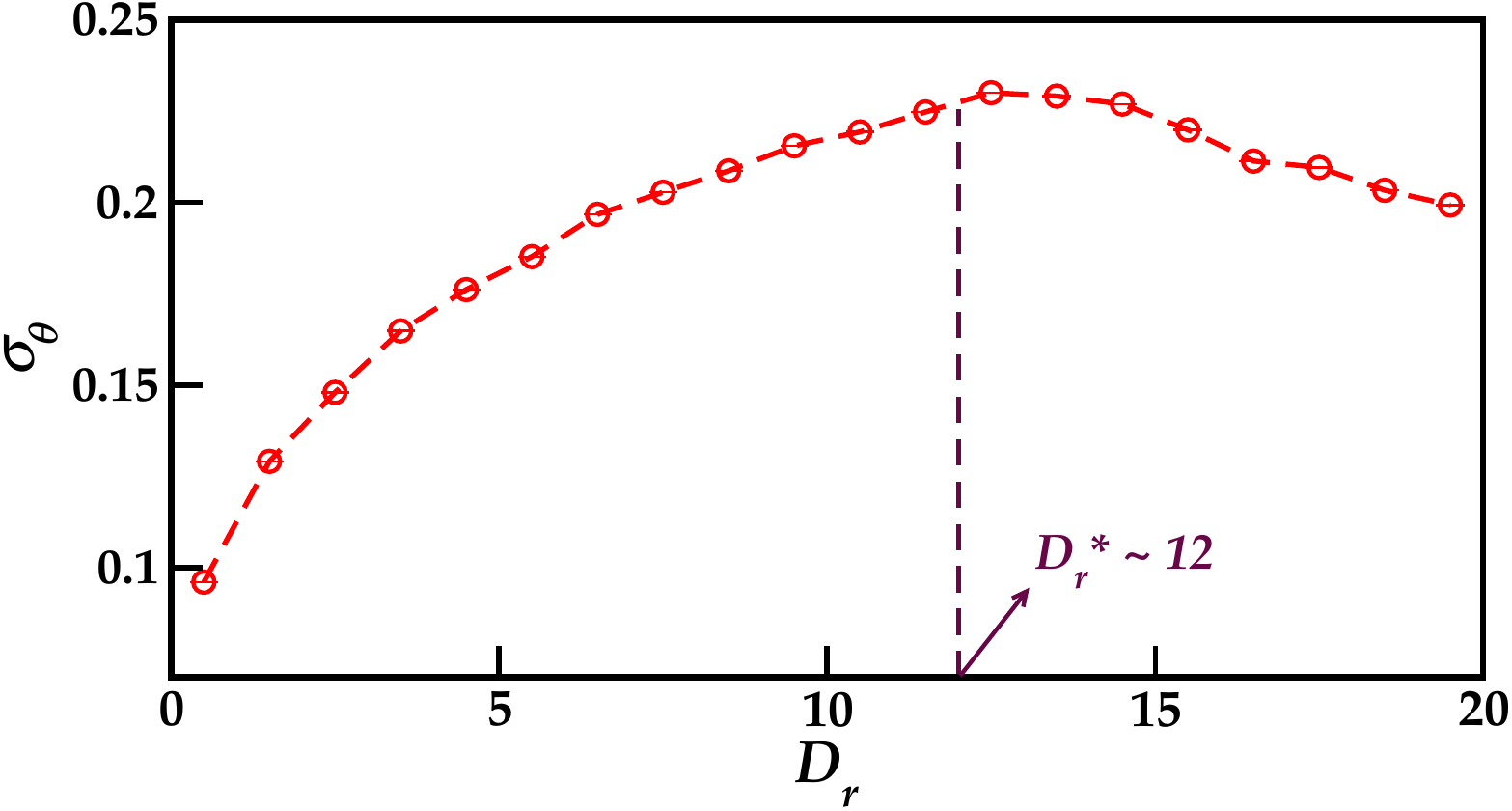}
    \caption{(Color online) The plot shows the variation of the standard deviation of angular deviation, $\theta(t)$, 
as a function of $D_r$ for the single-agent system showing narrowing of the angular distribution beyond the optimal $D_r^{*}$. Error bars are smaller than the size of the symbols.}
\label{fig: standard_delta}
\end{figure}

    \section{Standard Deviation $\sigma_{\theta}$ as a Function of $D_r$ : Multi Agent}
\label{standard_multi}    


To investigate why in a two-agents system one agent exhibits enhanced performance while the other becomes slower—even compared to the single-agent case—we analyze the standard deviation of the angular deviation $\theta(t)$ as a function of $D_r$, keeping all other simulation parameters identical. The results reveal that although the number of orientation resets increases with $D_r$, beyond the optimal value $D_r^{*}$ (of single particle) the distribution becomes progressively narrower, indicating reduced angular fluctuations. This trend suggests that frequent resetting leads to less orientational fluctuations for one particle, while constraining the adaptability of the other. 
Extending this analysis to a multi-particle system, we observe a consistent behavior: the fastest agent in a larger group exhibits an even smaller standard deviation in $\theta(t)$, confirming that cooperative interactions in larger assemblies further suppresses the orientation fluctuations and enhance the efficiency of the most adaptive agents.

\begin{figure}[htbp]
    \centering

    \begin{subfigure}[b]{0.8\textwidth}
        \centering        \includegraphics[width=\linewidth]{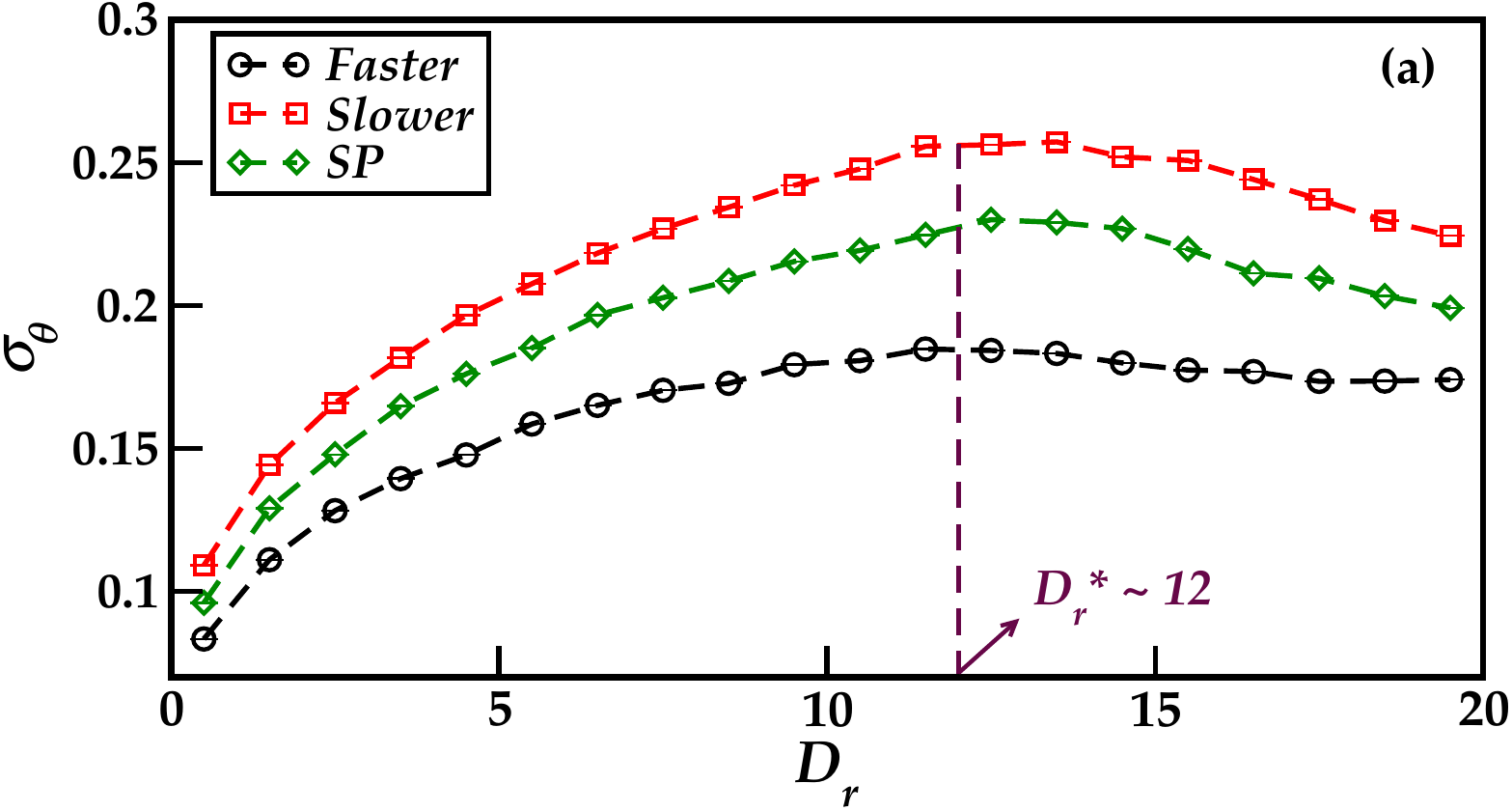}
        
        \label{fig:std_two_particle}
    \end{subfigure}
    \hfill
    \begin{subfigure}[b]{0.8\textwidth}
        \centering
        \includegraphics[width=\linewidth]{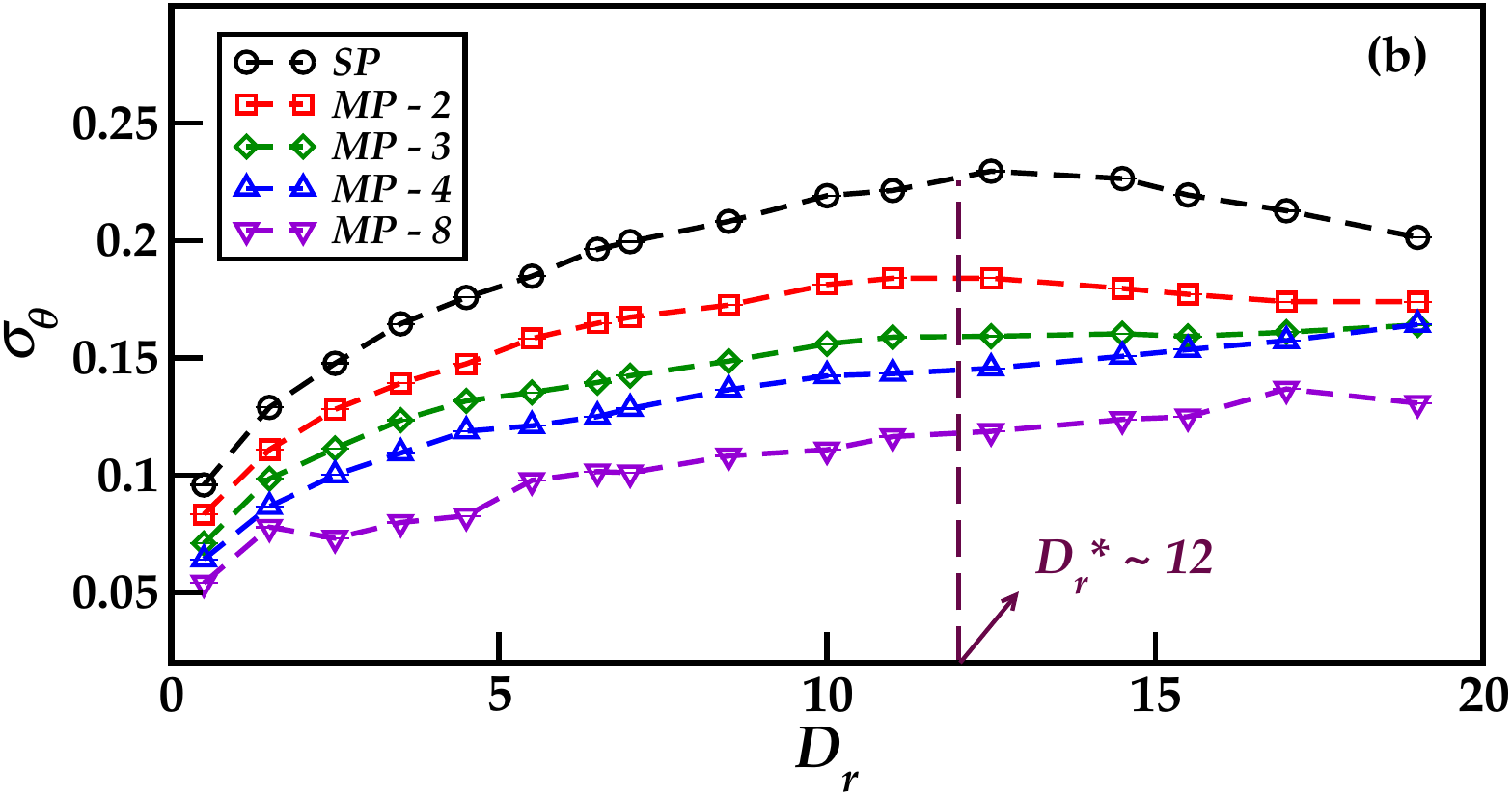}
        
        \label{fig:std_multi_particle}
    \end{subfigure}

    \caption{(Color online)Standard deviation of the angular deviation $\theta(t)$ $vs.$ $D_r$ for (a) the two-agents system and (b) multi-agents system (fastest particle), compared with the single-agent case. Results are averaged over 2000 realizations; error bars are smaller than the symbol size. 
In (a), SP denotes the single-particle case, while faster and slower indicate the two agents. In (b), MP--$N$ represents the fastest particle in an $N$-particle system ($N=2,3,4,8$).}
    \label{fig:std_theta_comparison}
\end{figure}
\clearpage

\section {Movies of two-agent dynamics and comparison between RL, ABP with resetting, and pure ABP at different $D_r$}.

\label{movies}
\large\textbf{Movie 1:}
Movie showing the motion of a two-agents system at $D_r = 15$. The RL agents start from nearby locations having  the same radial distance from home.\\
\textit{Link}:\url{https://drive.google.com/file/d/1D2iczKqoHdMWGplj3ozjNkVlGXq8c-8D/view?usp=sharing}
\\

\large\textbf{Movie 2:}
Movie showing the motion of a reinforcement-learning (RL) agent and an active Brownian particle (ABP) with resetting, at $D_r = 2$. The RL agent and the ABP start from different initial positions but at the same radial distance from home.\\
\textit{Link}:\url{https://drive.google.com/file/d/1ICVnwpck9ewT14tKIPIyl-tPMJBnyo4w/view?usp=sharing}
\\

\large\textbf{Movie 3:}
Movie showing the motion of a reinforcement-learning (RL) agent and an active Brownian particle (ABP) with resetting, at $D_r = 12$. The RL agent and the ABP start from different initial positions but at the same radial distance from home.\\
\textit{Link}:\url{https://drive.google.com/file/d/1KvLVZfynIBZ-mhTwd8XgJPIi6aQjn_1i/view?usp=sharing}
\\

\large\textbf{Movie 4:}
Movie showing the motion of a reinforcement-learning (RL) agent and an active Brownian particle (ABP) with resetting, at $D_r = 15$. The RL agent and the ABP start from different initial positions but at the same radial distance from home.\\
\textit{Link}:\url{https://drive.google.com/file/d/1np6BmJLBpzqgAYELSNIfAzM_d8_qT68L/view?usp=sharing}
\\

\large\textbf{Movie 5:}
Movie showing the motion of a reinforcement-learning (RL) agent, an active Brownian particle (ABP) with resetting and a pure ABP without resetting, at $D_r = 2$. All particles start from different initial positions but at the same radial distance from home.\\
\textit{Link}:\url{https://drive.google.com/file/d/1zTevDJ7doyOPpOlZ-pZ0gNpXstGU4jT8/view?usp=sharing}
\\

\twocolumngrid
\bibliography{citation}
\clearpage

\end{document}